\documentclass[lettersize,journal]{IEEEtran}
\usepackage{algorithmicx}
\usepackage{algpseudocode}
\usepackage{amsmath,amsfonts}
\usepackage{algorithm}
\usepackage{array}
\usepackage[caption=false,font=normalsize,labelfont=sf,textfont=sf]{subfig}
\usepackage{textcomp}
\usepackage{stfloats}
\usepackage{url}
\usepackage{verbatim}

\usepackage{lipsum}
\usepackage{multicol}
\usepackage{bm}
\usepackage{etoolbox}

\usepackage{times}
\usepackage{epsfig}
\usepackage{graphicx}
\usepackage{amsmath}
\usepackage{amssymb}

\usepackage{tabulary}
\usepackage[table]{xcolor} 
\usepackage{paralist}
\usepackage{booktabs}
\usepackage{adjustbox}
\usepackage{xspace}

\usepackage{epsfig}
\usepackage{graphicx}
\usepackage{cite}
\usepackage{multirow}
\newcommand{\PAR}[1]{\noindent{\bf #1}}

\usepackage{xcolor}

\definecolor{rblue}{rgb}{0,0.5,1}
\definecolor{awesome}{rgb}{1.0, 0.13, 0.32}
\definecolor{gy}{rgb}{0.5, 0, 0.5}
\def\eg{\textit{e.g.}} 
\def\ie{\textit{i.e}.} 

\def\etc{\textit{etc}.}

\definecolor{hollywoodcerise}{rgb}{0.96, 0.0, 0.63}
\definecolor{lasallegreen}{rgb}{0.03, 0.47, 0.19}
\definecolor{hanpurple}{rgb}{0.32, 0.09, 0.98}
\definecolor{green(pigment)}{rgb}{0.0, 0.65, 0.31}
\usepackage[pagebackref=true,breaklinks=true,colorlinks,bookmarks=false]{hyperref}
\hypersetup{colorlinks=true,linkcolor={red},citecolor={hanpurple},urlcolor={magenta}}

\begin{document}

\title{Exploring Quasi-Global Solutions to Compound Lens Based Computational Imaging Systems}

\author{Yao Gao$^{1}$, Qi Jiang$^{1}$, Shaohua Gao$^{1}$, Lei Sun$^{1}$, Kailun Yang$^{2,3}$, and Kaiwei Wang$^{1,}$\IEEEauthorrefmark{2}
\thanks{This work was supported in part by the National Natural Science Foundation of China (NSFC) under Grant No. 12174341 and No. 62473139 and in part by Hangzhou SurImage Technology Company Ltd.}%
\thanks{$^{1}$Y. Gao, Q. Jiang, S. Gao, L. Sun, and K. Wang are with the State Key Laboratory of Extreme Photonics and Instrumentation, Zhejiang University, Hangzhou 310027, China (e-mail: gaoyao@zju.edu.cn; qijiang@zju.edu.cn; gaoshaohua@zju.edu.cn; leo\_sun@zju.edu.cn; wangkaiwei@zju.edu.cn).}%
\thanks{$^{2}$K. Yang is with the School of Robotics, Hunan University, Changsha 410012, China (E-mail: kailun.yang@hnu.edu.cn).}%
\thanks{$^{3}$K. Yang is also with the National Engineering Research Center of Robot Visual Perception and Control Technology, Hunan University, Changsha 410082, China.}%
\thanks{\IEEEauthorrefmark{2}Corresponding author: Kaiwei Wang.}%
}

\markboth{IEEE Transactions on Computational Imaging, February~2025}%
{Gao \MakeLowercase{\textit{et al.}}: Quasi-Global Search Optics}

\IEEEpubid{}

\maketitle

\begin{abstract}
Recently, joint design approaches that simultaneously optimize optical systems and downstream algorithms through data-driven learning have demonstrated superior performance over traditional separate design approaches. However, current joint design approaches heavily rely on the manual identification of initial lenses,  posing challenges and limitations, particularly for compound lens systems with multiple potential starting points. In this work, we present Quasi-Global Search Optics (QGSO) to automatically design compound lens based computational imaging systems through two parts: (i) Fused Optimization Method for Automatic Optical Design (OptiFusion), which searches for diverse initial optical systems under certain design specifications; and (ii) Efficient Physic-aware Joint Optimization (EPJO), which conducts parallel joint optimization of initial optical systems and image reconstruction networks with the consideration of physical constraints, culminating in the selection of the optimal solution in all search results. Extensive experimental results illustrate that QGSO serves as a transformative end-to-end lens design paradigm for superior global search ability, which automatically provides compound lens based computational imaging systems with higher imaging quality compared to existing paradigms. The source code will be made publicly available at \url{https://github.com/LiGpy/QGSO}.
\end{abstract}


\section{Introduction}
\begin{figure}[!t]
  \centering
  \includegraphics[width=0.9\linewidth]{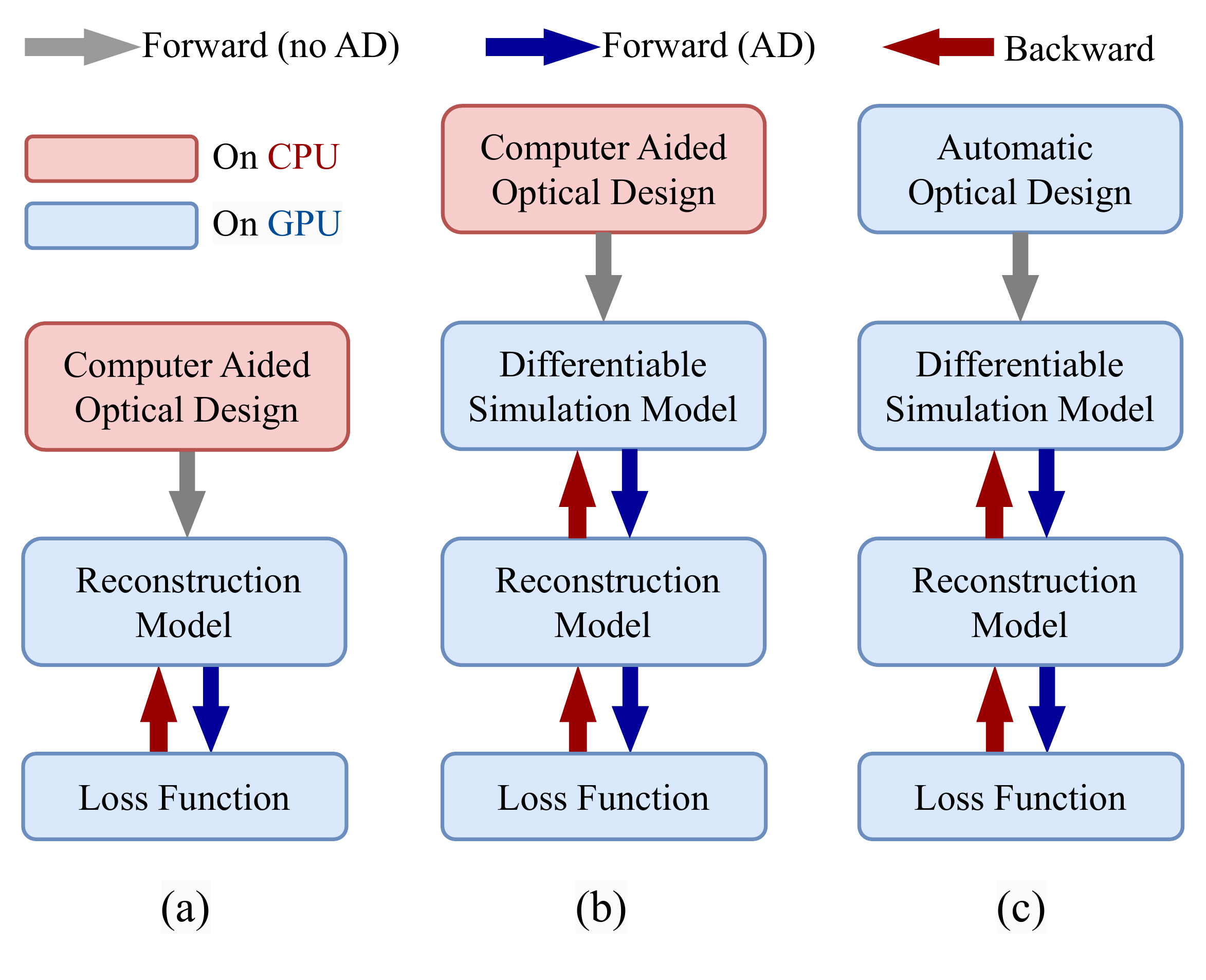}
  \caption{Comparison of the design modes for computational imaging systems. (a) shows the separate, sequential design mode. (b) shows the joint design mode that requires manual determination of the initial optical systems. (c) shows the proposed QGSO paradigm (joint design mode in which the algorithm automatically provides the initial optical systems).}
  \label{fig:compare}
\end{figure}

We are heading to a new era of mobile vision, in which more correction tasks are shifted from traditional optical design to image reconstruction algorithms, a process central to computational imaging~\cite{suo2023computational}.
Traditionally, the optical system and the image reconstruction model in computational imaging have been designed sequentially and separately, as depicted in Fig.~\ref{fig:compare}(a), which may not achieve the best cooperation of the two components~\cite{sitzmann2018end}.
Recent years have seen the rise of joint design pipelines that effectively bridge the gap between optical design and algorithmic development~\cite{sitzmann2018end,sun2021end,wang2022differentiable}. These paradigms leverage differentiable imaging simulation models within an automatic differentiation (AD) framework, enabling the joint optimization of optical systems and image reconstruction models. 

This paradigm has been widely applied successfully to the design of single-element optical systems, \eg, Diffractive Optical Element (DOE) or metasurface~\cite{sitzmann2018end,tseng2021neural,chang2019deep}. Furthermore, considerable efforts have been made to expand the paradigm to compound optical systems composed of multiple refractive optical elements~\cite{sun2021end,wang2022differentiable,tseng2021differentiable} and further expand the optimization variables to the full set of lens parameters~\cite{cote2023differentiable,zhang2023large}. 
However, the design of compound lenses presents a significant challenge due to their highly non-convex nature, making it difficult to commence with a random set of parameters solely relying on local optimization algorithms. Typically, an initial design exhibiting basic functional performance is developed first. This initial design is then refined through a process of joint optimization to improve visual task performance by trading off imaging quality across different fields of view, wavelengths, and depths~\cite{sun2021end, wang2022differentiable, tseng2021differentiable, cote2023differentiable,zhang2023large}. 
As illustrated in Fig.~\ref{fig:compare}(b), even with the involvement of the image reconstruction network, the traditional method of manually restricting the overall design space based on optical metrics, \eg, Modulation Transfer Function (MTF), Point Spread Function (PSF), \etc, does not obviate the necessity for skilled personnel. 
Moreover, there may be multiple potential initial structures with different aberration characteristics for compound lenses, and traversing all the potential designs manually is time-consuming and impracticable~\cite{zhou2024revealing}.

To address this issue, some studies on joint optimization have proposed to start from randomly initialized configurations, leveraging curriculum learning to reduce dependence on an initial design~\cite{yang2022automatic, yang2023curriculum}. Nevertheless, these approaches primarily focus on the automated design of optical lenses and do not delve into the manufacturing constraints associated with optical systems, potentially leading to the limitations of the optimized results in practical applications. 
Some studies have also proposed a Deep Neural Network (DNN) framework to automatically and quickly infer lens design starting points tailored to the desired specifications~\cite{cote2019extrapolating,cote2021deep}, and the trained model acts as a backbone for a web application called LensNet. However, the model is confined to basic specifications like effective focal length, F-number, and field of view, without accommodating more complex physical constraints, \eg, glass thickness, air spacing, total track length, back focal length, \etc, and the specifications that can be considered are limited by the existing optical system database.

\begin{figure*}[!t]
  \centering
  \includegraphics[width=1\linewidth]{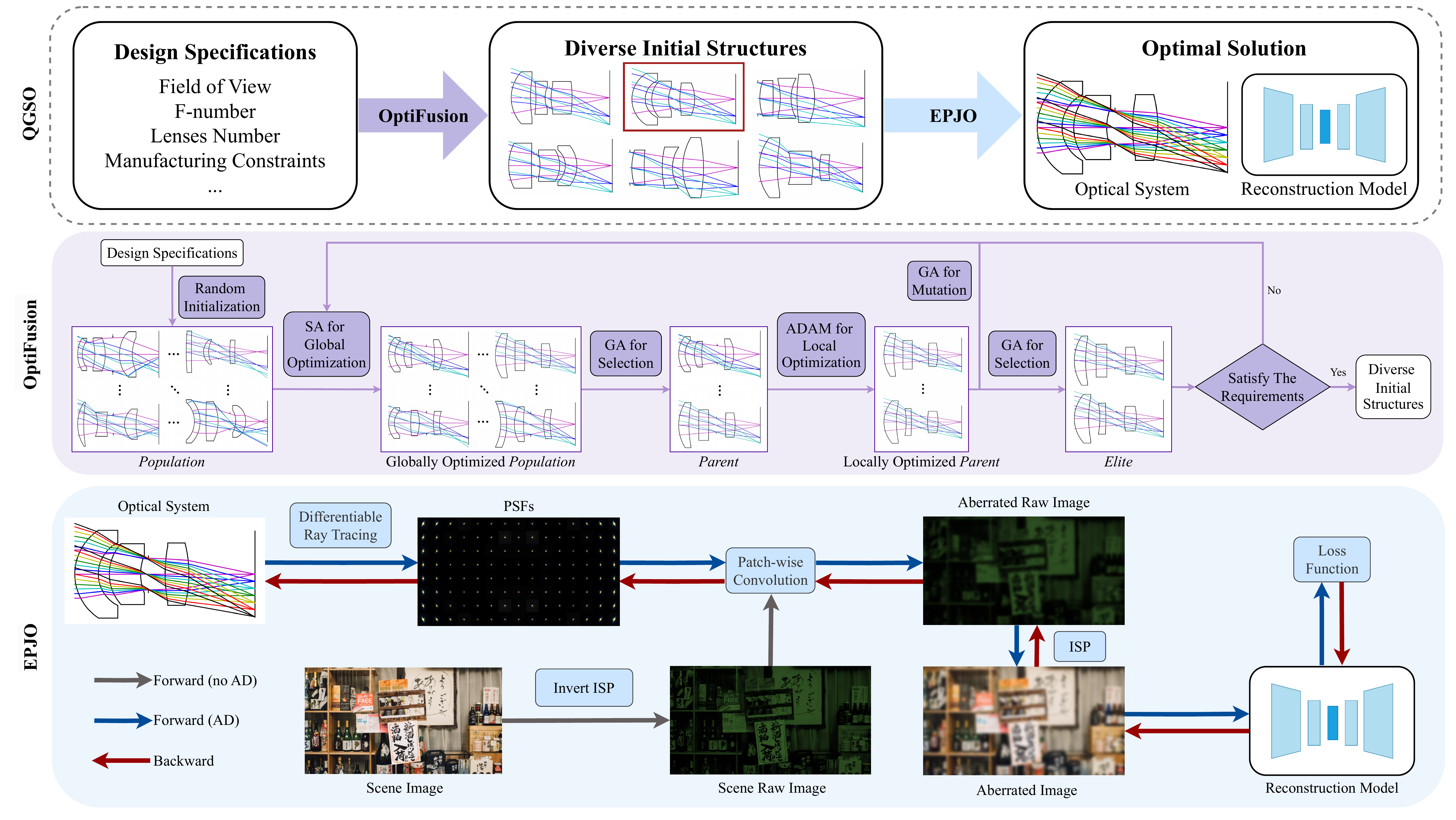}
  \vspace{-1.5em}
  \caption{Overview of our compound lens based computational imaging systems design method. Quasi-Global Search Optics (QGSO) includes the Fused Optimization Method for Automatic Optical Design (OptiFusion) and the Efficient Physic-aware Joint Optimization (EPJO). OptiFusion fuses Simulated Annealing (SA), Genetic Algorithm (GA), and ADAM to automatically search for initial structures with sufficient diversity based on traditional optical design metrics. EPJO includes an enhanced differentiable simulation model that incorporates differentiable ray tracing, patch-wise convolution, and an Image Signal Processing (ISP) pipeline. 
  Additionally, EPJO incorporates customized memory-efficient techniques that enable parallel joint optimization of the initial structures discovered by OptiFusion and image reconstruction models, within reasonable computational resources. This approach allows us to select the jointly optimal solution in all search results based on the final reconstructed image quality metrics.
  }
  \label{fig:overview}
  \vspace{-0.5em}
\end{figure*}

In this work, we introduce Quasi-Global Search Optics (QGSO), a comprehensive end-to-end lens design framework as shown in Fig.~\ref{fig:compare}(c), which bypasses the requirement for manual initial setting determination and features robust global search capabilities. For the sake of design efficiency and performance, we believe that establishing sound initial structures based on traditional optical design metrics remains essential for our joint design approach. 
Uniquely, QGSO includes the Fused Optimization Method for Automatic Optical Design (OptiFusion), which combines Simulated Annealing (SA), Genetic Algorithm (GA), and ADAM to autonomously find initial structures with adequate diversity rooted in traditional optical design metrics. QGSO also includes Efficient Physic-aware Joint Optimization (EPJO), featuring an advanced differentiable simulation model and customized memory-efficient techniques. This allows for parallel joint optimization of initial structures identified by OptiFusion and image reconstruction networks, efficiently using computational resources to select the optimal solution in all search results based on the final image quality metrics. Furthermore, EPJO considers more complex physical constraints of optical systems compared to existing works~\cite{cote2023differentiable, yang2023curriculum} and the categorical nature of glass materials to strongly encourage manufacturable outcomes. The overview of QGSO is shown in Fig.~\ref{fig:overview}.

The experimental results demonstrate that OptiFusion can traverse more diverse and reasonable initial designs compared to existing methods in multiple design forms such as Cooke Triplets or Double Gauss lenses. 
Then we highlight QGSO's enhanced global search capability by the end-to-end design of extended depth-of-field (EDoF) three-element lenses, illustrating a marked improvement over both joint design method that requires manual identification of initial structures and traditional separate and the sequential design method.
To summarize, our key contributions are:
\begin{itemize}
    \item Introduction of Quasi-Global Search Optics (QGSO), a comprehensive end-to-end lens design framework that thoroughly and autonomously explores the solution space for compound lens based computational imaging systems under certain design specifications.
    \item Validation of OptiFusion’s superior ability to search for diverse and suitable initial structures compared to existing automatic optical design methods.
    \item Validation of QGSO’s superior global search capability through comparison with both joint design method that requires manual determination of initial optical systems and the traditional separate design method.
\end{itemize}

\section{Related Work}
\subsection{Computational Imaging}
The aberration-induced image blur is inevitable due to insufficient lens groups for aberration correction~\cite{2011Modeling,mahajan1994zernike}. 
To this intent, computational imaging methods~\cite{2011Non,2013High} appear as a preferred solution, where optical designs with necessary optical components are equipped with an image reconstruction model. 
Early efforts have been made to solve the inverse image reconstruction problem through model-based methods~\cite{Schuler2012Blind,2015Blind}. Recently, learning-based methods~\cite{peng2019learned,chen2021extreme_quality, Chenshiqi, chen_mobile_2023, jiang2022annular, jiang2023minimalist, jiang2024realworld, jiang2022computational, zhou2024revealing, luo2024correcting} have been widely explored for delivering more impressive results of computational imaging, which benefits from the blooming development of image restoration~\cite{liang2021swinir,chen2022simple}, image super-resolution~\cite{wang2018esrgan,zhang2018image,liang2021swinir} and image deblur~\cite{zamir2022restormer,chen2022simple,wang2022uformer} methods. 
Further research has developed deep learning frameworks for the joint optimization of optical systems and reconstruction models, aiming to perfectly align them and thus enhance overall imaging performance~\cite{sitzmann2018end, li2021end, sun2021end, yang2023curriculum, yang2023design_compact}. Traditionally, joint design has relied on manually crafted lenses as initial points~\cite{sitzmann2018end, li2021end, sun2021end} or employed strategies like curriculum learning~\cite{yang2023curriculum} for optimizing random initial lenses and reconstruction models, somewhat restricting the breadth of global search capability.
Considering these limitations, this work introduces Quasi-Global Search Optics (QGSO), a novel framework for the design of compound lens based computational imaging systems, to automatically generate a variety of starting points for joint optimization, and efficiently achieve joint optimization of all starting points and reconstruction models. 

\subsection{Automatic Optical Design}
In the field of joint optimization of optical systems and post-processing models, generating a variety of initial optical system structures is essential. This need highlights the importance of automatic optical design, which seeks to develop algorithms capable of minimizing or even eliminating manual intervention in the design process. The Damped Least Squares (DLS) method, introduced by Kenneth Levenberg~\cite{levenberg1944method}, has been favored in engineering for its rapid convergence. 
However, DLS often becomes trapped in local minima, and it requires considerable expertise to establish a robust initial structure, limiting the potential for full automation. Efforts have been made to automate the inference of lens design starting points using Deep Neural Networks (DNN) tailored to specific requirements~\cite{cote2019extrapolating, cote2021deep}. Yet, the lack of a comprehensive optical system database restricts the diversity of the outputs, and the model is confined to basic specifications like effective focal length, F-number, and field of view, without accommodating more complex physical constraints. As algorithms and computational power have evolved, various heuristic global search algorithms, \eg, Simulated Annealing (SA), Genetic Algorithm (GA), Ant Colony Algorithm (ACA), Particle Swarm Optimization (PSO), and Tabu Search (TS), have become prevalent in automatic optical design~\cite{guo2019new, sun2021automatic, zhang2020automated, yue2022adaptive, tang2019ant, reichert2020development}. Nevertheless, the purpose of the above works is still to automatically design the optimal optical system under traditional design metrics, and the diversity of design results cannot be guaranteed. Consequently, we propose the Fused Optimization Method for Automatic Optical Design (OptiFusion), which combines Simulated Annealing (SA), Genetic Algorithm (GA), and ADAM to automatically search for diverse initial structures.

\subsection{Joint Optimization of Optical Systems and Image Processing Algorithms}
Due to the spatial variation in optical aberrations, which cannot be avoided during the lens design process, recent imaging systems have shifted some of these correction tasks from optical design to image processing algorithms~\cite{chen2021optical}. However, imaging systems have long been designed in separate steps: experience-driven optical design followed by sophisticated image processing~\cite{sun2021end}.
The joint optimization of optical systems and image processing algorithms represents a groundbreaking paradigm that has gained traction in recent years~\cite{sitzmann2018end, wetzstein2020inference, Sun_Tseng_Fu_Heidrich_Heide_2020, sun2021end, wang2022differentiable, tseng2021neural}. This paradigm has been applied successfully to the design of single-element optical systems composed of a single Diffractive Optical Element (DOE) or metasurface~\cite{sitzmann2018end, jeon2019compact, dun2020learned, baek2021single, chugunov2021mask, tseng2021neural, li2022quantization, bacca2021deep_coded} and has also been applied to the design of hybrid systems composed of an idealized thin lens combined with a DOE as an encoding element~\cite{chang2019deep, Sun_Tseng_Fu_Heidrich_Heide_2020, sun2020end, metzler2020deep, ikoma2021depth,shi2022seeing,pinilla2022hybrid}. 

More recently, efforts have been made to expand the paradigm to compound optical systems composed of multiple refractive optical elements~\cite{sun2021end, wang2022differentiable, tseng2021differentiable, cote2021deep, fontbonne2022comparison, yang2023curriculum, cote2023differentiable, yang2023image, zhang2023large}. 
However, many of these studies have neglected the intricate physical constraints inherent in real-world applications of optical systems~\cite{sun2021end,wang2022differentiable,tseng2021differentiable, yang2023image, fontbonne2022comparison}. 
Some works have merely imposed basic constraints, like ray angle~\cite{yang2023curriculum, cote2023differentiable}, which do not adequately address manufacturability concerns. 
Further, the substantial computational memory required for joint optimization continues to be a significant challenge, with some researchers questioning the feasibility of fully optimizing with the available computational resources~\cite{chen_mobile_2023, zhou2024revealing, zhang2023large}. 

This work proposes Efficient Physic-aware Joint Optimization (EPJO) to address these challenges. 
EPJO not only takes into account more complex physical constraints of optical systems and the categorical nature of glass materials to enhance their manufacturability but also achieves efficient joint optimization through customized memory-efficient techniques.

\section{OptiFusion: Proposed Method for Automatic Optical Design}
\label{OF}
OptiFusion is an evolutionary algorithm designed to automatically generate diverse initial optical systems for subsequent joint optimization. This method combines Genetic Algorithms (GA), Simulated Annealing (SA), and ADAM to optimize spot size and meet physical constraints (Sec. \ref{sec_OptiFusionA_1}). The foundational concept of OptiFusion is based on evolutionary theory, where all optical systems constitute a $Population$, and each system within is considered an $Individual$. OptiFusion begins with random initialization of the $Population$ (Sec.~\ref{sec_OptiFusionA_2}). Throughout the evolutionary process, each generation applies SA for preliminary global optimization (Sec.~\ref{sec_OptiFusionA_3}), followed by selection of a subset of the globally optimized $Population$ as the $Parent$, using GA’s selection mechanism (Sec.~\ref{sec_OptiFusionA_4}). ADAM then performs local optimization on the $Parent$ (Sec.~\ref{sec_OptiFusionA_5}). A select portion of this locally optimized $Parent$ group is designated as $Elite$. Should the evolutionary process continue, the $Parent$ undergoes mutation and is merged with the $Elite$ for further optimization in the subsequent generation (Sec.~\ref{sec_OptiFusionA_6}). If the evolutionary process concludes, the $Elite$ is finalized as the output. The specific procedures are outlined in Algorithm~\ref{algorithm:1}.

\begin{algorithm}[t!] 
\caption{Implementation steps of OptiFusion}
\label{algorithm:1}
\begin{algorithmic}[1]
\Require Design specifications, number of generations in GA ($N$), number of $Individuals$ ($m$)
\Ensure  The last generation of $Elite$ ($Z_{N}$)  

\State $X_{1}$ $\leftarrow$ Initialization() \Comment{Random Initialization}
    \For {$g = 1,2,...,N$} 
        \State $X'_{g}$ $\leftarrow$ SA($X_{g}$) \Comment{Global Optimization}
        \State $Y_{g}$ $\leftarrow$ SelectParent($X'_{g}$) \Comment{Select $Parent$}
        \State $Y'_{g}$ $\leftarrow$ ADAM($Y_{g}$) \Comment{Local Optimization}
        \State $Z_{g}$ $\leftarrow$ SelectElite($Y'_{g}$) \Comment{Select $Elite$}
        \State $M_{g}$ $\leftarrow$ Mutate($Y'_{g}$) \Comment{Mutate $Parent$}
        \State $X_{g+1}$ $\leftarrow$ Merge($M_{g}$, $Z_{g}$) \Comment{Next Generation}
                    
    \EndFor
    \State \Return $Z_{N}$ 
    
\end{algorithmic} 
\end{algorithm}

\subsection{Loss Function of OptiFusion}
\label{sec_OptiFusionA_1}
OptiFusion models a compound lens as a stack of several spherical glass elements, characterized by their curvatures ($c$), glass and air spacings ($s$), and the refractive indexes ($n$) and Abbe numbers ($v$) at the ``d'' Fraunhofer line ($587.6nm$). 
Following~\cite{sun2021end}, we employ the approximate dispersion model $n(\lambda) \approx A + B/\lambda^2$ to retrieve the refractive index at any wavelength $\lambda$, where $A$ and $B$ follow the definition of the ``d''-line refractive index and Abbe number. 
Once the field of view and aperture size are set, ensuring no vignetting occurs, we express the normalized lens parameters — including curvatures, spacings, refractive indexes, and Abbe numbers — as an $n$-dimensional vector
\begin{equation}
\label{eq:OptiFusion_loss_0}
\bm{x}=(x^{(1)}, x^{(2)},...,x^{(n)})^T \in \mathbb{R}^n.
\end{equation}
Here, all lens parameters are optimized variables. In addition, all variables are normalized according to their corresponding value ranges, allowing for unified operations on different types of variables.
The primary objective is to optimize $\bm{x}$ to minimize a specific loss function $\mathcal{L}(\bm{x})$. 
Conventional lens design tasks usually seek a design of suitable complexity that fulfills a given list of specifications; these are translated into a loss function that targets optical performance criteria as well as many manufacturing constraints~\cite{cote2023differentiable}. Therefore, the specific loss function of OptiFusion needs to include imaging quality loss and physical constraint loss simultaneously.

\PAR{Imaging Quality Loss.}
Traditional lens designs focus on straightforward metrics such as the basic aberrations, spot RMS radius, or MTF. Spot RMS radius and basic aberrations such as chromatic aberration are easier to calculate compared to MTF and are more suitable for evaluating systems with large aberrations.
In OptiFusion, therefore, to expedite the search for viable initial structures, we integrate a spot loss ($\mathcal{L}_S$) and a lateral chromatic aberration loss ($\mathcal{L}_{LC}$) to assess an optical system:
\begin{equation}
\label{eq:OptiFusion_loss_1}
\mathcal{L}_{IQ}=\mathcal{L}_S+\alpha_{LC}\mathcal{L}_{LC}.
\end{equation}
Here, $\mathcal{L}_S$ quantifies the average spot RMS radius across all sampled fields of view and wavelengths~\cite{Foreman:74}. And $\mathcal{L}_{LC}$ accounts for the average lateral chromatic aberration~\cite{mahajan1991aberration}. Please refer to the Appendix for detailed definitions of $\mathcal{L}_S$ and $\mathcal{L}_{LC}$. We typically set $\alpha_{LC}$ at 0.25 to maintain an optimal balance. 

\PAR{Physical Constraint Loss.}
For basic parameters $\bm{x}$, we straightforwardly constrain their normalized values within the range $[0,1]$. However, for key physical properties, \eg, effective focal length and total track length, which are derived from $\bm{x}$, it's imperative to incorporate a soft physical constraint loss ($\mathcal{L}_{PC}$) to align with design specifications using the Lagrangian approach. Suppose there are $n_i$ physical quantities to be constrained, with each quantity $q_i$ subject to a lower threshold $q^{(i)}_{min}$ and an upper threshold $q^{(i)}_{max}$, along with a specified weight $\alpha_i$. The $\mathcal{L}_{PC}$ is then expressed as:

\begin{equation}
\label{eq:OptiFusion_loss_2}
\mathcal{L}_{PC} = \dfrac{1}{n_i}\sum_{i}\alpha_i[\max(q^{(i)}_{min} - q_i, 0)+\max(q_i - q^{(i)}_{max} , 0)].
\end{equation}
This formulation implies a linear penalty for any deviation of $q_i$ from the interval $[q^{(i)}_{min}, q^{(i)}_{max}]$, ensuring effective constraints on physical quantities during the design process. In addition, this formulation including the max operator is differentiable by using numerical differentiation~\cite{cote2023differentiable}, so it is applicable to both global optimization algorithms and local optimization algorithms.

\PAR{OptiFusion Loss.}
The overall loss function for OptiFusion, denoted as $\mathcal{L}_{OF}$, is formulated as:
\begin{equation}
\label{eq:OptiFusion_loss_3}
\mathcal{L}_{OF} = \mathcal{L}_{PC} + \alpha_{IQ}\mathcal{L}_{IQ}.
\end{equation}
Here, $\alpha_{IQ}$ is set to $1$, balancing the emphasis on imaging quality with other design considerations, such as effective focal length and total track length. And the unit of $\mathcal{L}_{OF}$ is millimeters. When multiple working object distances are specified in the design, the average value of $\mathcal{L}_{OF}$ across all distances serves as the aggregate loss for optimization purposes.

\subsection{Initialization}
\label{sec_OptiFusionA_2}
To reduce reliance on manual input from optical designers and enable fully automated design, OptiFusion begins with the random initialization of the $Population$, which comprises $m$ $Individuals$ expressed as:

\begin{equation}
\label{eq:OptiFusion_ini_1}
X= \{ \bm{x}_1, \bm{x}_2 ,..., \bm{x}_{m} \}.
\end{equation}
A larger value of $m$ theoretically means faster global search as more optical systems are optimized in parallel. Although a larger $m$ value improves the performance, it also results in higher memory consumption and is limited by the computing power of the GPU. Therefore, it is necessary to choose a reasonable $m$ value based on the computing device.
Each $\bm{x}_i$ is a randomly initialized individual, structured as per Eq.~(\ref{eq:OptiFusion_loss_0}). In terms of the parameters, normalized curvatures and spacings within $\bm{x}_i$ are randomized within the range $[0,1]$. Differently, the normalized refractive indexes and Abbe numbers are set to either $0$ or $1$, based on established optical design insights suggesting that extreme values of refractive indexes and Abbe numbers often enhance the imaging performance of simple optical systems.

\subsection{SA for Preliminary Global Optimization of Population}
\label{sec_OptiFusionA_3}
Simulated Annealing (SA) is a heuristic algorithm that mimics the thermodynamic process of cooling to achieve global optimization by potentially accepting suboptimal solutions to escape local minima. Unlike gradient-based methods such as ADAM or DLS, SA does not require derivative information, thereby reducing computational demands. Thus, SA is particularly suited for a preliminary global search when facing a large number of highly inferior $Individuals$ to be optimized.

SA iteratively optimizes $Population$. 
During each iteration, assuming that $\forall \bm{x}_i \in X $, we calculate the loss $\mathcal{L}_{i}$ based on Eq.~(\ref{eq:OptiFusion_loss_3}) and adjust the annealing temperature to improve adaptability:

\begin{equation}
\label{eq:sa_1}
T_{i}=\alpha_{SA}\mathcal{L}_{i},
\end{equation}

where $\alpha_{SA}$ is predefined as $0.1$. A random perturbation $\Delta \bm{x}_{i} \in (-0.1, 0.1)$ is applied to $\bm{x}_{i}$, yielding a new $Individual$ $\bm{x}'_{i}$ and its loss $\mathcal{L}'_{i}$. The change in loss, $\Delta \mathcal{L}_{i}=\mathcal{L}'_{i}-\mathcal{L}_{i}$, determines the acceptance probability of $\bm{x}'_{i}$:
\begin{equation}
\label{eq:sa_2}
P_{i}=\min(e^{-\Delta \mathcal{L}_{i} / T_{i}},1).
\end{equation}
A random number $\epsilon\in (0, 1)$ is drawn; if $\epsilon{<}P_{i}$, $\bm{x}_{i}$ is updated to $\bm{x}'_{i}$; otherwise, it remains unchanged. Furthermore, SA tracks the best historical solution and its loss $\mathcal{L}^{best}_{i}$ for each $Individual$, utilizing this information to gauge the progress towards convergence. In general, we define the mean loss value of $Population$ as
\begin{equation}
\label{eq:sa_3}
\mathcal{L}_{mean}=\frac{1}{m} \sum^m_{i=1} \mathcal{L}^{best}_{i}.
\end{equation}
When the rate of decrease of $\mathcal{L}_{mean}$ is less than the threshold, which is typically set to $0.025$, it is considered that the global optimization has reached convergence, and we output the set of historical optimal $Individuals$ for further selection:
\begin{equation}
\label{eq:sa_4}
X'= \{ \bm{x}^{best}_1, \bm{x}^{best}_2 ,..., \bm{x}^{best}_{m} \}.
\end{equation}

\subsection{Selection of Parent}
\label{sec_OptiFusionA_4}
The $Parent$ is selected as a subset of $X'$, denoted as $Y$:

\begin{equation}
\label{eq:sel_1}
Y= \{ \bm{y}_1, \bm{y}_2 ,..., \bm{y}_{m'} \} \subset X',
\end{equation}
where $m'{=}r(0.06\cdot m)$ and $r(\cdot)$ represents rounding to the nearest integer. The parameter value $0.06$ represents the proportion of excellent lenses selected from the globally optimized lens group. Because local optimization requires greater computational power than global optimization, as gradients need to be calculated, a small number of lenses need to be selected from the globally optimized lenses for subsequent local optimization. The larger the value of this parameter, the better, as more optical systems can be selected for subsequent local optimization, providing more possible structures. Due to the limitation of the device's computing power, however, the value of this parameter is empirically set to $0.06$. To curate a collection of high-quality and diverse $Y$ from $X'$, we refine the Genetic Algorithm's (GA) selection process to better suit optical design. We begin by defining:
\begin{equation}
\label{eq:sel_2}
\mathcal{L}_{all}= \{ \mathcal{L}^{best}_{1}, \mathcal{L}^{best}_{2} ,..., \mathcal{L}^{best}_{m} \}.
\end{equation}
We then sort $X'$ based on $\mathcal{L}_{all}$ and select $Y$ from $X'$ prioritizing from highest to lowest quality. To prevent the selection of overly similar optical systems and maintain diversity within the $Parent$ group, we measure the Euclidean distance $d{=}\|\bm{x}'_i-\bm{x}'_j\|$ between ${\forall} \bm{x}'_i,\bm{x}'_j{\in}X'$. 
If $d{\leq}0.2$, only the superior individual is chosen for inclusion in $Y$. Through this process, there is a small batch of good optical systems, \ie, $Parent$, to be selected from the current lens group, \ie, $Population$.

\subsection{ADAM for Local Optimization of Parent}
\label{sec_OptiFusionA_5}
Despite the quick convergence offered by the Damped Least Squares (DLS) method, its computational speed can significantly decrease as the number of variables increases, due to the necessity for matrix inversion. Alternatively, ADAM~\cite{kingma2014adam}, known for its efficient local optimization and adaptive learning rate adjustments, is more apt for automatic optical design. ADAM does require gradient information for parameter optimization; however, in cases of the relatively simple $\mathcal{L}_{OF}$, effective optimization can be achieved using the first-order difference quotient as a gradient approximation. 
This approach avoids the need for differentiable simulation models and substantially reduces memory usage.

Thus, we employ ADAM to optimize the $Parent$ group $Y$, selected as Sec.~\ref{sec_OptiFusionA_4}, towards local optima. We also implement a cosine annealing learning rate schedule to enhance the robustness of ADAM's optimization process. The optimization steps and convergence criteria align with those described in Sec.~\ref{sec_OptiFusionA_3}, leading to the optimization of the $Parent$, now denoted as $Y'$.

\subsection{Selection of Elite and Mutation of Parent}
\label{sec_OptiFusionA_6}
The $Elite$ group is selected from the subset of $Y'$ and is denoted as $Z$:
\begin{equation}
\label{eq:mut_1}
Z= \{ \bm{z}_1, \bm{z}_2 ,..., \bm{z}_{m''} \} \subset Y',
\end{equation}
where $m''{=}r(0.02\cdot m)$. 
This selection process follows the mechanisms outlined in Sec.~\ref{sec_OptiFusionA_4}. If the process has surpassed the predetermined number of generations, the $Elite$ becomes the final output; otherwise, it is carried over to the next generation to ensure that high-quality optical systems are not discarded through the evolutionary process. Additionally, to expand the exploration of potential solutions for subsequent generations, mutation operations are applied to $Y'$. $\forall \bm{y}'_i \in Y'$, a number $n_{mut}$ of variables are randomly altered within the range $[0,1]$, where $n_{mut}$ is set to $r(0.3\cdot n)$. 
Moreover, the total length of the optical system is kept constant pre- and post-mutation to ensure the rationality of the mutation results. The mutated results, represented as $M$, along with the $Elite$ $Z$, are then merged to form the $Population$ $X$ for the next generation.

\section{EPJO: Proposed Pipeline for Joint Optimization}
\label{EPJO}
This section outlines the differential imaging simulation model presented in Sec.~\ref{EPJO_1}, which facilitates simultaneous optimization of the optical system and the image reconstruction network. Sec.~\ref{EPJO_2} defines the loss function of EPJO. Then we introduce a customized adjoint back-propagation strategy for memory-efficient in Sec.~\ref{EPJO_3}. Finally, we described the detailed steps of EPJO for joint optimization in Sec.~\ref{EPJO_4}.

\subsection{Differentiable Imaging Simulation Model}
\label{EPJO_1}
We establish an accurate differentiable simulation model suitable for compound optical systems under dominant geometrical aberrations, which achieves gradient back-propagation from image reconstruction network parameters to optical system parameters.

\PAR{Differentiable PSF Formation Model.} 
In our differentiable imaging simulation pipeline, the aberration-induced degradation is represented through the energy dispersion of the Point Spread Function (PSF).
We employ a ray-tracing-based model for PSF formation that enables accurate and differentiable results. Differentiable ray tracing is achieved by alternating between updating the coordinates of the rays from one interface to the next using the Newton iteration method and updating the direction cosines following Snell's Law as in~\cite{sun2021end} and~\cite{wang2022differentiable}. 
Rays are initially positioned at the entrance pupil, and a ray-aiming correction step~\cite{cote2023differentiable} is applied to ensure precise simulation of optical systems, particularly those affected by pupil aberrations. 
Then, rays can be traced to the image plane to obtain the PSF. Under dominant geometrical aberrations, diffraction can be safely ignored \cite{cote2023differentiable}, and the PSF can be calculated by Gaussianizing the intersection of the ray with the image plane~\cite{li2021end}. Specifically, assuming the number of traced rays is $n_{ray}$, at specific sampled fields of view $\theta$ and sampled wavelengths $\lambda_c$, the PSF is composed of $t{\times}t$ pixels and the center of the PSF is set as the intersection of the chief ray and the image plane as in~\cite{chen2021optical}. And then the PSF can be modeled as:
\begin{equation}
\label{eq:simu1}
    \small PSF(\theta, \lambda_c)=
    \Bigg\{ \frac{1}{\sqrt{2\pi}\sigma}\sum_{k=1}^{n_{ray}}\exp(-\frac{d_{i,j}^k(\theta, \lambda_c)^2}{2\sigma^2}) \Bigg\}_{\substack{1\leq i \leq t \\ 1\leq j \leq t}}.
\end{equation}
Here, $(i, j)$ is the index of the pixel of the PSF, $k$ is the index of the traced rays, 
$d_{i,j}^k(\theta, \lambda_c)$ represents the distance between the pixel $(i,j)$ and the intersection of the $k_{th}$ ray with the image plane, and $\sigma{=}\sqrt{\Delta x^2{+}\Delta y^2}{/}3$, in which $\Delta x{\times}\Delta y$ is physical size of each pixel, so the pixel closest to the intersection can obtain $99.7\%$ of the intensity proportion. Please refer to~\cite{li2021end} for more details.

After obtaining PSFs for all sampled fields and wavelengths using the aforementioned methods, we synthesize them into three-channel RGB PSFs. This synthesis utilizes the spectral sensitivity characteristics of the simulated CMOS sensor, as follows:
\begin{equation}
\label{eq:simu2}
    PSF_c(\theta) = \sum_{\lambda_c}W_c(\lambda_c) \cdot PSF(\theta, \lambda_c).
\end{equation}
Here, $\theta$ represents the sampled fields of view, and $c$ represents one of R, G, and B channels. $\lambda_c$ represents sampling wavelengths of the corresponding channel and $W_c(\lambda_c)$ represents the corresponding normalized wavelength response coefficient. 
Moreover, it is essential to account for the influence of longitudinal chromatic aberration on the central positioning of each channel within the three-channel RGB PSFs. 
Therefore, we designate the center of the G-channel PSF as the reference point for the RGB PSFs, adjusting the PSFs of the R and B channels based on their actual central positions. Consequently, we generate the integrated three-channel RGB PSFs across all sampled fields of view.

\PAR{Patch-wise Convolution and ISP Pipeline.} 
To facilitate the construction of more realistic aberrated images, an Image Signal Processing (ISP) pipeline is employed~\cite{brooks2019unprocessing}. Initially, the scene image $I_S$ undergoes sequential applications of inverse Gamma Correction (GC), inverse Color Correction Matrix (CCM), and inverse White Balance (WB) to transform it into the scene raw image $I'_S$. The inverse ISP pipeline is expressed as:
\begin{equation}
\label{eq:simu3}
    I'_S = P^{-1}_{WB} \circ P^{-1}_{CCM} \circ P^{-1}_{GC}(I_S),
\end{equation}
where $\circ$ denotes the composition operator, and $P_{WB}$, $P_{CCM}$, and $P_{GC}$ represent the procedures for WB, CCM, and GC, respectively. 

Subsequently, patch-wise convolution is applied to $I'_S$. $I'_S$ is divided into $n_h{\times}n_w$ patches, each measuring $s{\times}s$ pixels. It is assumed that PSFs within these patches are spatially uniform. Convolution is then performed between the image patches and their corresponding PSFs, which are then recompiled into the degraded raw image $I'_A$. Each patch of $I'_S$ is designated as $I'_S(c, h, w)$, where $c$ indicates one of the R, G, and B channels, and $h$ and $w$ denote the patch’s position on the image plane. The associated PSF, $PSF (c, h, w)$, is derived by interpolating PSFs across all sampled fields of view and adjusting them by rotating to the correct angle:
\begin{equation}
\label{eq:simu4}
    PSF(c, h, w) = P_{rot}(\sum_{\theta} W(\theta)\cdot PSF(c, \theta)),
\end{equation}
where $P_{rot}$ represents the rotation process, and rotation angle $\beta$ can be calculated by $\beta = \arctan(w/h)$. $PSF(c, \theta)$  is the PSF from a specific field of view and $W(\theta)$ is the normalized interpolation weight determined by the inverse square formula. The degraded raw image patch $I'_A(c, h, w)$ is approximated as:
\begin{equation}
\label{eq:simu5}
    I'_A(c, h, w) \approx PSF(c, h, w) \ast I'_S(c, h, w).
\end{equation}
After that, we stitch $n_h{\times}n_w$ degraded raw image patches to obtain the complete degraded raw image $I'_A$ in the same order as we previously split $I'_S$, and then we mosaic the degraded raw image $I'_A$ before adding shot and read noise to each channel. Moreover, we sequentially apply the demosaic algorithm, WB, CCM, and GC to the R-G-G-B noisy raw image, where the aberration-degraded image $I_A$ in the sRGB domain is obtained. The ISP pipeline can be defined as:
\begin{equation}
\label{eq:simu6}
\begin{aligned}
    I_A = P_{GC} \circ P_{CCM} \circ P_{WB} \circ  P_{demosaic} \circ  (P_{mosaic}(I'_A) + N),
\end{aligned}
\end{equation}
where $N$ represents the Gaussian shot and read noise. $P_{mosaic}$ and $P_{demosaic}$ represent the procedures of mosaicking and demosaicking, respectively.

\PAR{Discussion.}
Overall, the entire differentiable simulation model includes three parts: Differentiable PSF Formation Model, Patch-wise Convolution, and ISP Pipeline. Although some existing works~\cite{li2021end, cote2023differentiable, chen2021optical} have demonstrated the accuracy of this simulation model to some extent, there may still be small discrepancies between the simulation model and the real imaging results because it essentially approximates and simplifies the real imaging process. In the future, with the application of more advanced simulation models, it is potential to gradually narrow the gap between simulation and real imaging.

\subsection{Loss Function of EPJO}
\label{EPJO_2}
We define the loss function of EPJO balancing the emphasis on final reconstructed image quality with consideration of intricate physical constraints.

\PAR{Imaging Quality Loss.}
We reconstruct aberration-degraded images $I_A$ through an image reconstruction network to produce reconstructed images $I_R$. To extend the depth of field in compound lens based computational imaging systems, we segment the continuous object distance range into three training depths. The imaging quality loss function is formulated as:
\begin{equation}
\begin{aligned}
\label{eq:loss1}
    \mathcal{L}'_{IQ}=\frac{1}{3}\sum_{j} [\mathcal{L}_{mse}(I_{Rj}, I_{S}) + \alpha_{1} \mathcal{L}_{perc}(I_{Rj}, I_{S})] \\ + \sum_{j \neq 2}\alpha_{2}\mathcal{L}_{mse}(I_{Rj}, I_{R2}),
\end{aligned}
\end{equation}
where $\mathcal{L}_{mse}$ denotes the MSE loss, and $\mathcal{L}_{perc}$ indicates the perceptual loss function based on the pre-trained VGG16 network~\cite{zhang2018unreasonable}, enhancing alignment with human perception. And $\mathcal{L}_{mse}(I_{Rj}, 
I_{R2})$ means that we take $I_{R2}$ as a reference to keep reconstructed images depth-invariant. We set $\alpha_{1}{=}0.01$, $\alpha_{2}{=}0.1$.

\PAR{Physical Constraint Loss.}
Our joint optimization process, EPJO, also imposes strict constraints on relevant physical quantities and aligns glass variables with catalog glasses to ensure manufacturability. 
The physical constraint loss function is given by:
\begin{equation}
\begin{aligned}
\label{eq:loss2}
\mathcal{L}'_{PC} = \dfrac{1}{n_i}\sum_{i}\alpha_i[(\max(q^{(i)}_{min} - q_i, 0) \\ +\max(q_i - q^{(i)}_{max} , 0))]^2  + \mathcal{L}_{gv},
\end{aligned}
\end{equation}
where $\mathcal{L}_{GV}$ minimizes the squared distance between each set of continuous glass variables and the nearest catalog glass:
\begin{equation}
\begin{aligned}
\label{eq:loss3}
\mathcal{L}_{GV} = \frac{1}{p} \sum^{p}_{i=1}\min (\alpha_n {\|}n_i-\bm{n_{cat}}{\|}^2_2 + \alpha_v {\|}v_i-\bm{v_{cat}}{\|}^2_2),
\end{aligned}
\end{equation}
where $p$ is number of lenses and empirically we set $\alpha_n{=}100$, $\alpha_v{=}0.0004$. 
Unlike Eq.~(\ref{eq:OptiFusion_loss_2}), Eq.~(\ref{eq:loss2}) implies a more severe quadratic penalty instead of a linear penalty for any deviation of $q_i$ from the interval $[q^{(i)}_{min}, q^{(i)}_{max}]$, which is more suitable for optical systems that have already roughly met the specifications.

\PAR{EPJO Loss.}
To balance imaging quality and physical constraints effectively, we define the EPJO loss as:
\begin{equation}
\label{eq:loss4}
\mathcal{L}_{EPJO} = \mathcal{L}'_{PC} + \alpha'_{IQ}\mathcal{L}'_{IQ},
\end{equation}
in which $\alpha'_{IQ}$ is empirically set to $100$. 

\subsection{Adjoint Back-propagation for Memory Savings}
\label{EPJO_3}
When the loss function is in the image space (\eg~Eq.~(\ref{eq:loss1})) which involves calculating a large number of PSFs, simulating high-resolution images, and going through image reconstruction networks, straightforward back-propagation requires unaffordable device memory.
The work of~\cite{wang2022differentiable} has proposed an adjoint back-propagation approach that splits forward computations into multiple passes to alleviate the back-propagation memory issue. Unfortunately, our differentiable imaging simulation model is based on the convolution of PSFs and images rather than relying on image rendering in which many millions of Monte Carlo rays are sampled~\cite{wang2022differentiable}, which makes existing adjoint methods not directly applicable. Therefore, we propose a customized adjoint back-propagation method for our differentiable imaging simulation model. 

\begin{algorithm}[t!]
\caption{Implementation steps of EPJO}
\label{algorithm:2}
\begin{algorithmic}[1]
    \Require Lenses number ($p$), initial optical system ($O$) and randomly initialized image reconstruction model ($R$)
    \Ensure  Jointly optimized optical system ($O'_{p}$) and image reconstruction model ($R'_{p}$)  
    \State \{$O'_0$, $R'_0$\} $\leftarrow$ JointOptimize(\{$O$, $R$\}) \Comment{Continuous Glass}
    \For {$j = 1,2,...,p$} 
    
        \State $O_j$ $\leftarrow$ SelectGlass($O'_{j-1}$, $j$) \Comment{Select Catalog Glass}
        \State $R_j$ $\leftarrow$ $R'_{j-1}$ 
        \State \{$O'_j$, $R'_j$\} $\leftarrow$ JointOptimize(\{$O_j$, $R_j$\}) 
                    
    \EndFor
    
    \State \Return \{$O'_{p}$, $R'_{p}$\}
    
\end{algorithmic} 
\end{algorithm}

Fundamentally, the device memory of our differentiable simulation model is mainly consumed in storing intermediate variables for calculating a large number of PSFs. 
Therefore, we propose a novel adjoint approach to manually separate the calculation of PSFs from subsequent steps. Given the loss function $\mathcal{L}_{EPJO}$, our goal is to evolve variable parameters $\xi$ iteratively towards an optimal $\xi'$ using gradient-based optimization, and this requires computing $\partial\mathcal{L}_{EPJO}/\partial\xi$, the partial derivatives that indicate how design parameters affect the error metric locally. Assuming $F(\xi)$ is a continuous function of $\xi$ for calculating PSFs, $\partial\mathcal{L}_{EPJO}/\partial\xi$ can be represented by the chain rule as:
\begin{equation}
\label{eq:adjoint1}
\frac{\partial\mathcal{L}_{EPJO}}{\partial\xi}=\frac{\partial\mathcal{L}_{EPJO}}{\partial F(\xi)} \frac{\partial F(\xi)}{\partial \xi} .
\end{equation}
According to Eq.~(\ref{eq:adjoint1}), after calculating PSFs, we perform the first back-propagation to obtain ${\partial F(\xi)}/{\partial \xi}$, the partial derivatives of PSFs with respect to the optical system parameters. Then, we store $F(\xi)$ and ${\partial F(\xi)}/{\partial \xi}$ while clearing the computation graph and intermediate variables because the memory consumption for storing $F(\xi)$ and ${\partial F(\xi)}/{\partial \xi}$ is much smaller.
Subsequently, we take PSFs as a differentiable input to calculate $\mathcal{L}_{EPJO}$. Finally, we conduct a second back-propagation to obtain $\partial\mathcal{L}_{EPJO} / \partial F(\xi)$, and thus we can obtain ${\partial\mathcal{L}_{EPJO}}/{\partial\xi}$ according to Eq.~(\ref{eq:adjoint1}). 
Since the computation time in joint optimization is mainly spent on the image reconstruction network rather than calculating PSFs, the additional time required to perform the first back-propagation to calculate ${\partial F(\xi)}/{\partial \xi}$ can be ignored. Therefore, such an adjoint back-propagation approach significantly reduces memory consumption to an affordable level without affecting the optimization time.

\subsection{Implementation Steps of Joint Optimization}
\label{EPJO_4}
Unlike training individual image reconstruction networks, joint optimization requires a focused approach that takes into account the distinct characteristics of both optical systems and networks. Therefore, we have tailored exclusive steps specifically for joint optimization, as outlined in Algorithm~\ref{algorithm:2}.

\PAR{Stopping Rules.}
In each epoch, the optimization process involves $n_{O}$ iterations for adjusting the optical system parameters. Within each iteration dedicated to the optical system, $n_{R}$ iterations are performed to fine-tune the network parameters, ensuring their adaptability to change in the optical system. 
After each iteration of optimizing the optical system, we evaluate its performance on the validation dataset. The combination of optical system and network parameters that yields the best performance among the $n_{O}$ iterations of each epoch is selected as the optimal configuration for that epoch. If the best performance of the current epoch fails to surpass the best performance of the previous epoch, it is considered that the joint optimization has reached a good and stopping state. We empirically set $n_{O}{=}5$, $n_{R}{=}1000$ to ensure the smooth progress of the entire optimization process.

\PAR{Replacing Continuous Glass Variables With Real Glass.}
Given the discrete nature of glass materials, our initial approach involves optimizing the refractive index and Abbe number of the material as continuous variables. Using Eq.~(\ref{eq:loss3}) as a guiding principle, we gradually move towards the realization of actual materials within the solution space.
Subsequently, to translate these continuous variables into the desired catalog glass material, we employ a step-wise substitution method. This involves systematically selecting the glass materials that require replacement in a prescribed order.
Once the computational imaging system is optimized to satisfy convergence conditions, we proceed to replace the chosen continuous variables with the closest matching material from our glass library. This replacement is based on Eq.~(\ref{eq:loss3}), and the corresponding variables are subsequently fixed. The process continues with retraining until convergence is achieved, followed by the replacement of the next glass component, until all glass materials have been replaced.

\section{Experiments And Results}
\label{exper}
In this section, we investigate the effectiveness of the proposed method through two experiments. Firstly, as a crucial component of QGSO, it is necessary to prove that OptiFusion can globally traverse possible initial designs in multiple design forms. In existing automatic optical design methods, LensNet~\cite{cote2021deep} can automatically and quickly infer lens design starting points tailored to the desired specifications, and we compare the proposed OptiFusion against LensNet in Sec.~\ref{sec_exper_1}. After investigating the ability of OptiFusion to automatically search for diverse initial designs, we need to further study the benefits of QGSO in improving the final performance of compound lens based computational imaging systems. Designing an Extended Depth-of-Field (EDoF) camera is challenging because it is complicated by the strong spatial variation of aberrations across the depths~\cite{yang2023curriculum}, which is suitable for evaluating the global search capability of QGSO. Therefore, we compare QGSO to existing paradigms through the end-to-end design of EDoF three-element lenses in Sec.~\ref{sec_exper_2}.

\subsection{Comparison Experiment Between OptiFusion and LensNet}
\label{sec_exper_1}
It should be noted that LensNet is confined to basic specifications including Effective Focal Length (EFL), F-number, and Half Field Of View (HFOV), without accommodating more complex physical constraints that can be considered by OptiFusion. 
Therefore, to ensure a fair comparison, we first set a certain design specification with a $40mm$ EFL, an F-number of $2.5$, and an HFOV of $20^{\circ}$ and use LensNet to produce designs under this specification. 
After obtaining the output results of LensNet in multiple design forms, we apply OptiFusion to produce designs with reasonable physical constraints that are consistent with LensNet in each design form. Please refer to the Appendix for a detailed setting of physical constraints corresponding to each design form.
The physical quantities that necessitate soft constraints included in $\mathcal{L}_{PC}$ (Eq.~(\ref{eq:OptiFusion_loss_2})) are EFL, distortion, air edge spacing, glass edge thickness, Back Focal Length (BFL), Total Track Length (TTL), and image height, and their respective weights $\alpha_{i}$ are set to $\{0.1,1,0.1,0.1,0.05,0.01,1\}$ according to their respective constraint ranges and optical design experience. 
In addition, $\mathcal{L}_{OF}$ (Eq.~(\ref{eq:OptiFusion_loss_3})) serves as the unifying evaluation metric for optical systems. 
We set the number of generations $N$ to $30$, the number of $Individuals$ $m$ to $4000$, and the $\mathcal{L}_{OF}$ of output lenses to be less than $0.04$ when utilizing OptiFusion.
To ensure the diversity of output results, the Euclidean distance between different optical systems output in the same design form is set to be no less than $0.25$. 
Please refer to Sec.~\ref{OF} for a detailed description of the entire process.

\begin{figure*}[!t]
  \centering
  \includegraphics[width=0.98\linewidth]{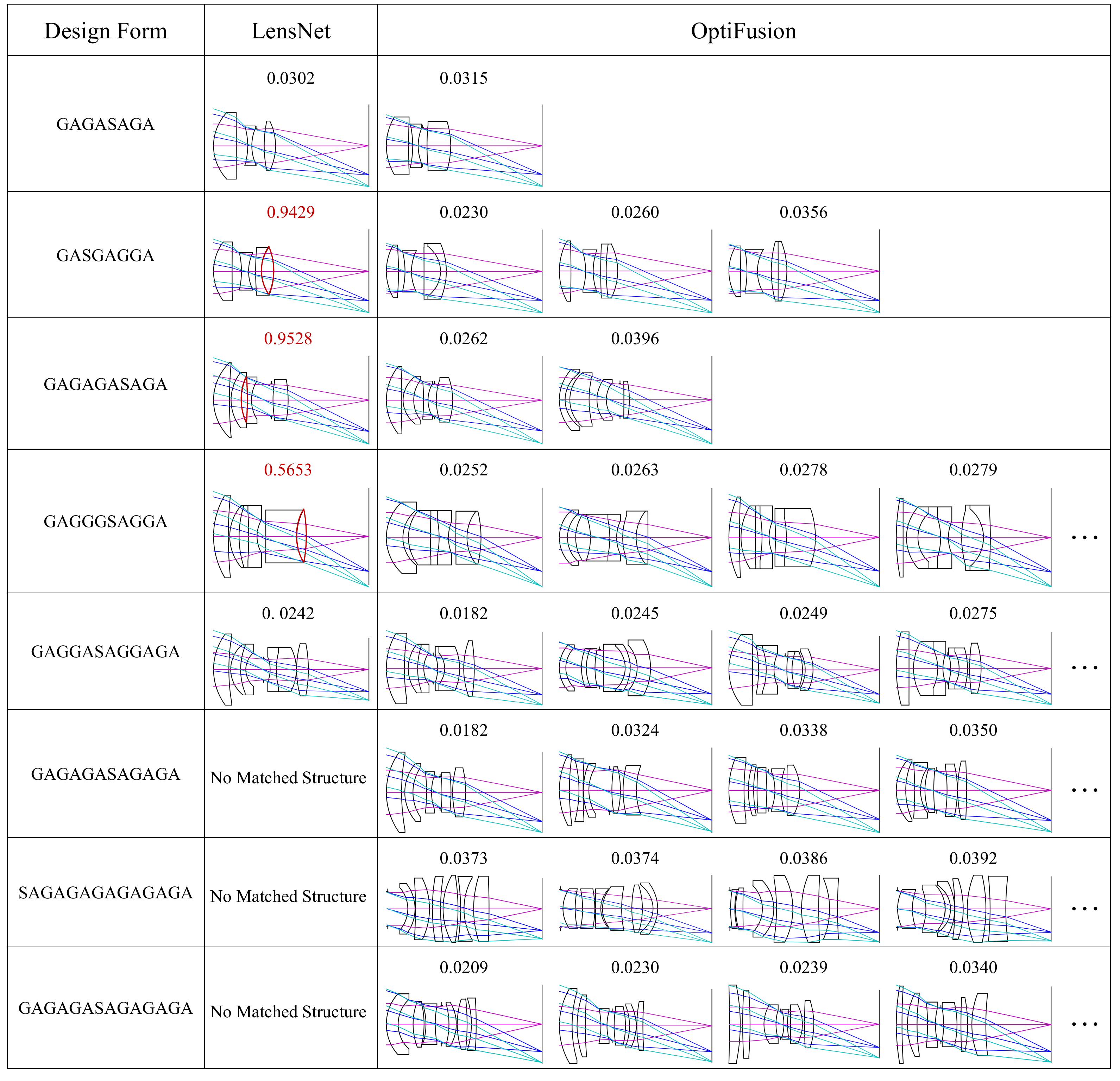}
  \vspace{-1.0em}
  \caption{Comparison between LensNet and OptiFusion in multiple design forms. The same as~\cite{cote2021deep}, each design form is named after their sequence of \textbf{G}lass elements, \textbf{A}ir gaps and aperture \textbf{S}top. The corresponding $\mathcal{L}_{OF}$ is marked above each optical system. LensNet may output the result with overlapping surfaces in some design forms (GASGAGGA, GAGAGASAGA, and GAGGGSAGGA), which makes $\mathcal{L}_{OF}$ abnormally large. The overlapping lens surfaces and the corresponding abnormally large $\mathcal{L}_{OF}$ are marked in red.}
  \label{fig:of_lensnet}
  \vspace{-1.5em}
\end{figure*}

\PAR{Experimental Results.}
From the perspective of design efficiency, the advantage of LensNet lies in its ability to quickly output design results (within one minute), as it is a trained network model. 
However, experiments have shown that within the complexity of designing six-element lenses, OptiFusion's design time can be reasonably controlled within $2$ hours, which is completely acceptable compared to the several days required for subsequent joint optimization of optical systems and image processing algorithms. More importantly, the design results indicate that OptiFusion has significant advantages over LensNet. Fig.~\ref{fig:of_lensnet} offers a visual comparison of the design outcomes in multiple design forms between LensNet and OptiFusion, and the corresponding $\mathcal{L}_{OF}$ is marked above each optical system. 
The same as~\cite{cote2021deep}, each design form is named after their sequence of \textbf{G}lass elements, \textbf{A}ir gaps and aperture \textbf{S}top. 
In addition to the design forms that LensNet can provide for design results, we also add three design forms (GAGAGASAGAGA, SAGAGAGAGAGAGA, and GAGAGASAGAGAGA) about five- or six-element lenses. 

When the design form is simple three-element and the position of the aperture stop is fixed between the second element and third element (GAGASAGA), LensNet and OptiFusion can both output the classic Cooke Triplets with similar $\mathcal{L}_{OF}$. 
However, when the design form becomes complex to four-element or six-element, LensNet can only provide up to one design in each certain design form, and there are even no matched structures in some other design forms (GAGAGASAGAGA, SAGAGAGAGAGAGA, and GAGAGASAGAGAGA), because it may be difficult for models trained through optical system databases to infer lenses in design forms not available in the optical system databases.
In contrast, OptiFusion can not only provide more than one lens in each design form but also handle more design forms than LensNet because it can perform a global search completely from scratch according to design requirements. 

Moreover, it should be noted that LensNet may output the result with overlapping surfaces in certain design forms (GASGAGGA, GAGAGASAGA, and GAGGGSAGGA), which is not in line with actual physical constraints and makes $\mathcal{L}_{OF}$ abnormally large. 
The overlapping lens surfaces and the corresponding abnormally large $\mathcal{L}_{OF}$ are marked in red in Fig.~\ref{fig:of_lensnet}. 
In contrast, due to the inclusion of corresponding physical constraints in the optimization objective, OptiFusion ensures that the design results strictly meet the given physical constraint requirements.

Overall, compared to LensNet, OptiFusion has the following advantages:
\begin{itemize}
    \item OptiFusion can meet more physical constraints specified by users, not just confined to EFL, F-number, and HFOV. OptiFusion ensures that the design results strictly meet the given physical constraint requirements.
    \item OptiFusion can search for more than one initial structure in a certain design form.
    \item OptiFusion is not limited by existing optical system databases and can handle more design forms within its design capabilities.
\end{itemize}

\subsection{End-to-end Design of EDoF Three-element Lenses}
\label{sec_exper_2}
We establish two representative specifications for three-element (3E) EDoF spherical lens designs, as outlined in Table~\ref{tab:design specifications}.
Specifically, 3E-\uppercase\expandafter{\romannumeral1} necessitates a HFOV of $20^{\circ}$ coupled with a F-number of $2.5$, whereas 3E-\uppercase\expandafter{\romannumeral2} requires a broader HFOV of $32^{\circ}$ and a smaller aperture with a F-number of $4.0$. 
We establish several distinct working distances (\textbf{D}epths) for each design specification similar to~\cite{sun2021end}. 
For each specification, we allow the position of the aperture to be variable, so OptiFusion performs the search for initial lenses simultaneously in four design forms (SAGAGAGA, GASAGAGA, GAGASAGA, and GAGAGASA). 
Please refer to Sec.~\ref{sec_exper_1} for detailed settings. Given the incorporation of heuristic random search algorithms, we employ OptiFusion to design each form three times to mitigate the effects of randomness. Afterward, the design results of all design forms are sorted according to $\mathcal{L}_{OF}$, and the top-$10$ are selected as the initial structures searched by OptiFusion.
Because LensNet cannot directly output results that meet all the design requirements in Table~\ref{tab:design specifications}, for comparison, we conduct initial lens design with the assistance of CODE V based on RMS spot size, \ie, the manual identification of lens design starting points, which we call the CODE V Assisted Joint Design (CAJD). 
Based on the above settings, for each design specification, QGSO provides $10$ diverse initial three-element lenses, whereas CAJD provides $1$ initial structure with the best aberration correction from the perspective of traditional optical design. 

Next, EPJO performs the same joint optimization on all initial structures, and detailed settings need to be determined.

\begin{table}[!t]
    \begin{center}
        \caption{Design specifications for two types of Three-element Lenses.}
        \vspace{-1.5em}
        \label{tab:design specifications}
        \resizebox{0.5\textwidth}{!}
{
\renewcommand{\arraystretch}{1.2}
\setlength{\tabcolsep}{1mm}{

\begin{tabular}{ccc}
\hline
Parameters  & 3E-\uppercase\expandafter{\romannumeral1}  & 3E-\uppercase\expandafter{\romannumeral2}  \\ \hline \hline
\multicolumn{1}{c|}{HFOV}          & $20^{\circ}$               & $32^{\circ}$               \\
\multicolumn{1}{c|}{F-number}               & $2.5$              & $4.0$               \\
\multicolumn{1}{c|}{EFL} & $38mm\sim42mm$    & $21mm\sim25mm$    \\
\multicolumn{1}{c|}{Working distance}       & $100m, 10m, 5m$     & $5m, 1m, 0.5m$      \\
\multicolumn{1}{c|}{Distortion}             & $-2\%\sim2\%$     & $-8\%\sim8\%$     \\
\hline
\multicolumn{1}{c|}{Curvature}              & \multicolumn{2}{c}{$-0.1\sim0.1$}  \\
\multicolumn{1}{c|}{Semi-diameter}          & \multicolumn{2}{c}{$<20mm$}  \\
\multicolumn{1}{c|}{Air center spacing}     & \multicolumn{2}{c}{$1mm\sim15mm$}  \\
\multicolumn{1}{c|}{Air edge spacing}       & \multicolumn{2}{c}{$1mm\sim15mm$}  \\ 
\multicolumn{1}{c|}{Glass center thickness} & \multicolumn{2}{c}{$4mm\sim15mm$}  \\
\multicolumn{1}{c|}{Glass edge thickness}   & \multicolumn{2}{c}{$5mm\sim15mm$}  \\
\multicolumn{1}{c|}{Refractive index}       & \multicolumn{2}{c}{$1.51\sim1.76$}  \\
\multicolumn{1}{c|}{Abbe number}            & \multicolumn{2}{c}{$27.5\sim71.3$}  \\
\multicolumn{1}{c|}{Wavelength}             & \multicolumn{2}{c}{$486nm, 588nm, 656nm$}  \\
\multicolumn{1}{c|}{BFL}  & \multicolumn{2}{c}{$>18mm$}  \\
\multicolumn{1}{c|}{TTL}           & \multicolumn{2}{c}{$<60mm$}  \\
\multicolumn{1}{c|}{Image height}  & \multicolumn{2}{c}{$14.16mm\sim14.44mm$}                     \\
\hline
\end{tabular}
}
}
    \end{center}
    \vspace{-2.0em}
\end{table}

\PAR{Differentiable Imaging Simulation.} 
To match the image height in Table~\ref{tab:design specifications}, we employ a virtual sensor with diagonal $d{=}28.6mm$. The sensor resolution is set to $1920{\times}1280$ pixels and the pixel size is $12.394{\mu}m$. 
We sample $5$ wavelengths for each channel based on a quantum efficiency curve that follows the Sony IMX172 sensor similar to~\cite{cote2023differentiable}. 
To reasonably control the speed and memory consumption of differentiable imaging simulation, we uniformly sample the PSF of $7$ fields of view and get the PSFs of the non-sampling field point by interpolation. We assume that the PSFs in the range of $64{\times}64$ pixels are spatially uniform, so every image is split into $30{\times}20$ patches that are $64{\times}64$ in size. 

\PAR{Data Preparation.} 
We adopt DIV2K~\cite{Martin_Fowlkes_Tal_Malik_2002} which contains $900$ images of $2K$ resolution as ground truths and divide these images into the training set and validation set at $8{:}1$. 
Then, images of different sizes are center-cropped and rotated to $1920{\times}1280$ pixels to match the sensor resolution, and images with length or width less than that of the sensor resolution will be discarded. 
Finally, we have obtained a training set containing $697$ images and a validation set containing $92$ images. 

\begin{figure}[!t]
  \centering
  \includegraphics[width=1.0\linewidth]{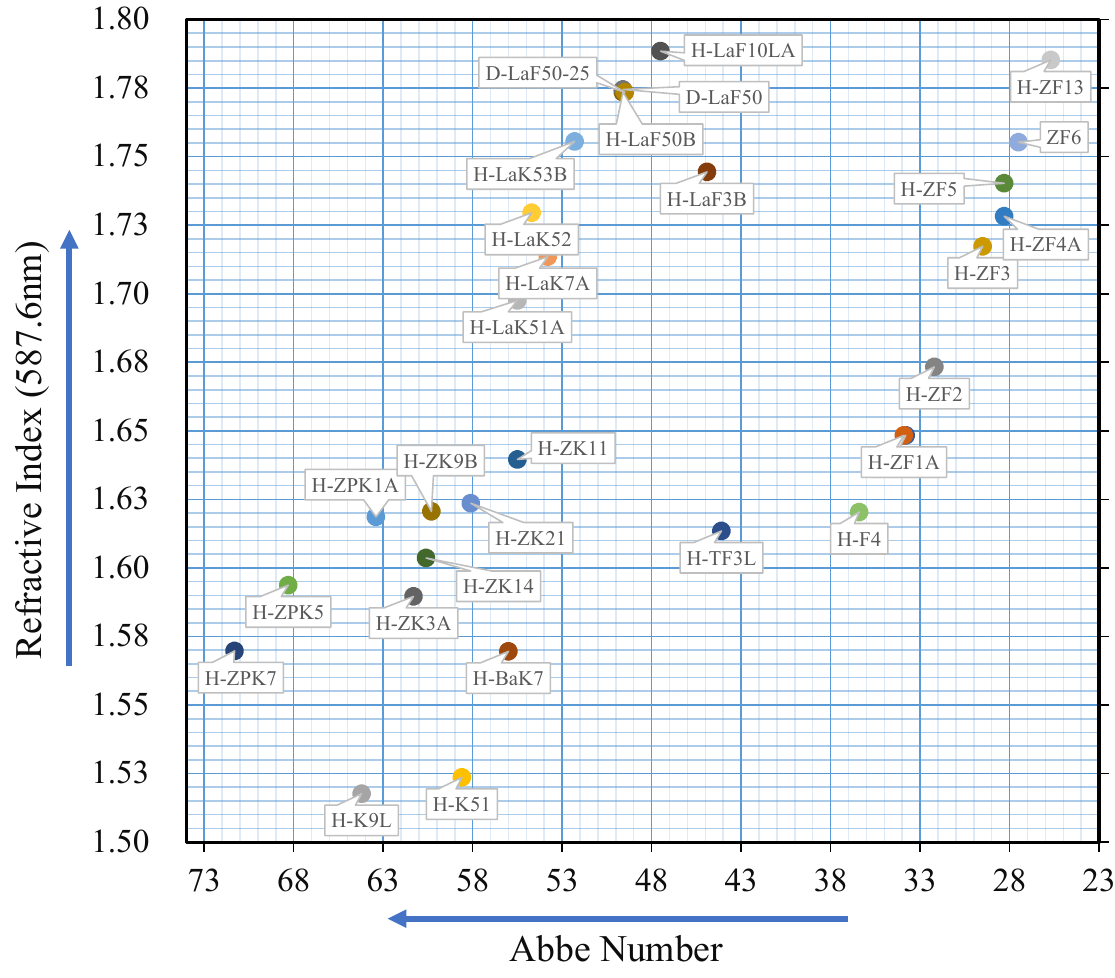}
  \caption{Catalog glasses that meet the design specifications and are available in stock all year round from the Chengdu Guangming Optoelectronic Corporation in China.}
  \label{fig:glass}
  \vspace{-1.5em}
\end{figure}

\PAR{Catalog Glasses.} 
To convert continuous glass variables into catalog glasses, we use glasses that meet the design specifications and are available in stock all year round from the Chengdu Guangming Optoelectronic Corporation in China, as shown in Fig.~\ref{fig:glass}.

\PAR{Training Details.} 
We use SwinIR~\cite{liang2021swinir} as the image reconstruction network without modifying the architecture. The Residual Swin Transformer Blocks (RSTB) number, Swin Transformer Layer (STL) number, window size, channel number, and attention head number are generally set to $5$, $6$, $8$, $96$, and $6$, respectively. 
During training, the patch size and batch size are set to $256{\times}256$ and $12$ respectively, in which each of the $3$ working distances occupies $4$ batch size. The ADAM optimizer with different learning rates is utilized, considering the respective characteristics of optical system variables and network variables. 
Specifically, the learning rates for curvature, spacing, refractive index, and Abbe number are set to $0.0002$, $0.02$, $0.001$, and $0.2$ respectively, and the learning rate of the network is $0.0001$. 
To achieve joint optimization of all initial lenses and reconstruction models, we implement EPJO in PyTorch~\cite{Paszke_Gross2019} on $22$ NVIDIA GeForce RTX 3090 GPUs for $32$ hours. 

\begin{figure*}[!t]
  \centering
  \includegraphics[width=1.0\linewidth]{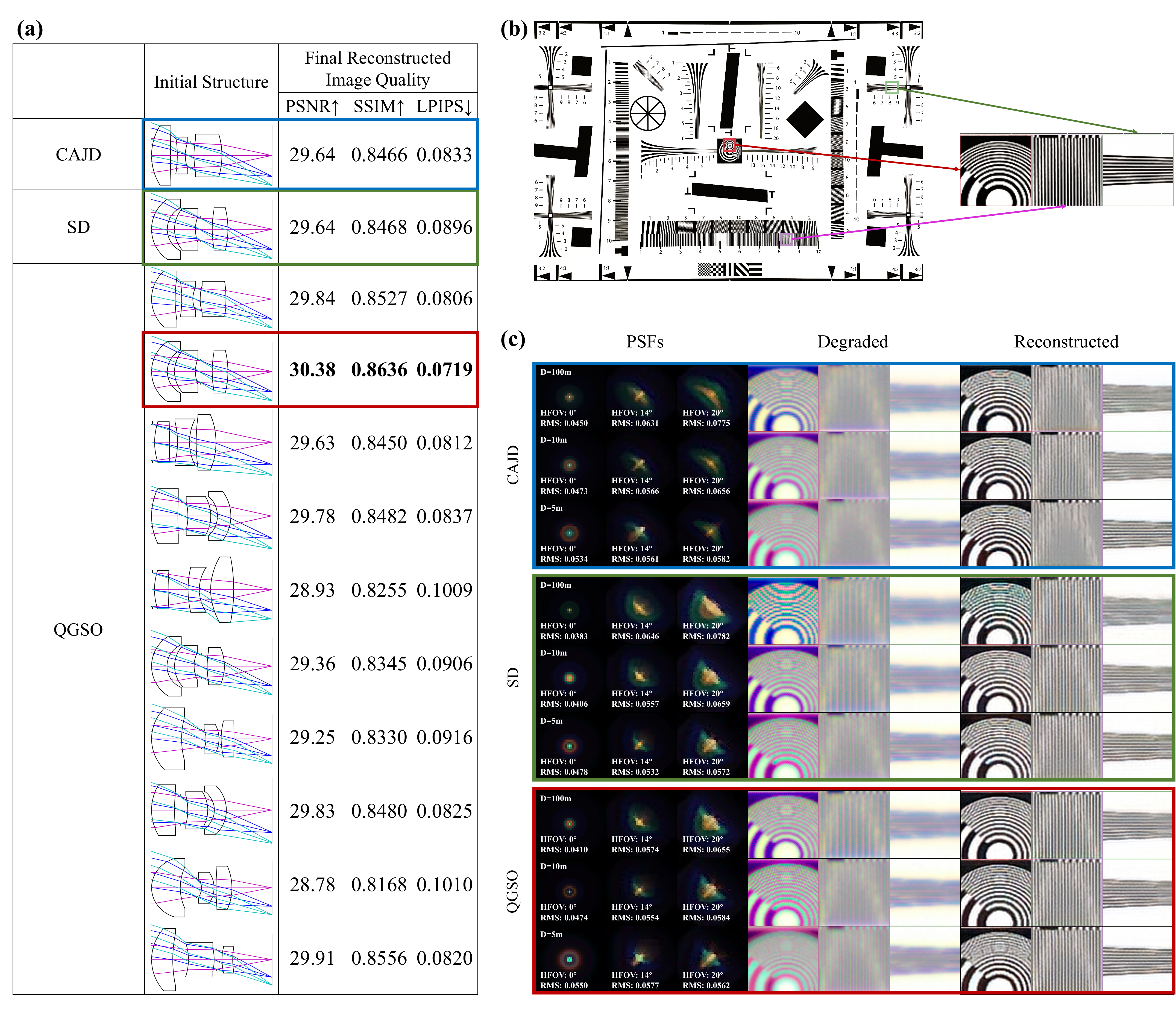}
  \vspace{-2.0em}
  \caption{Comparison between CAJD (CODE V Assisted Joint Design), SD (separate design), and QGSO under \textbf{3E-\uppercase\expandafter{\romannumeral1}}. (a) the initial structures and corresponding final reconstructed image quality of three methods. (b) the resolution chart (ISO 12233) taken by iPhone 12 and zoomed patches were used to evaluate image quality. (c) for each method and from left to right, we show 1) PSFs and corresponding RMS size ($mm$) across $3$ \textbf{D}epths (top: D${=}100m$; middle: D${=}10m$; bottom: D${=}5m$) and $3$ HFOV (left: $0^\circ$; middle: $14^\circ$; right: $20^\circ$); 2) degraded zoomed patches (top: D${=}100m$; middle: D${=}10m$; bottom: D${=}5m$); and 3) reconstructed zoomed patches (top: D${=}100m$; middle: D${=}10m$; bottom: D${=}5m$).}
  \label{fig:psf1}
  \vspace{-1.5em}
\end{figure*}

\begin{figure*}[!t]
  \centering
  \includegraphics[width=1.0\linewidth]{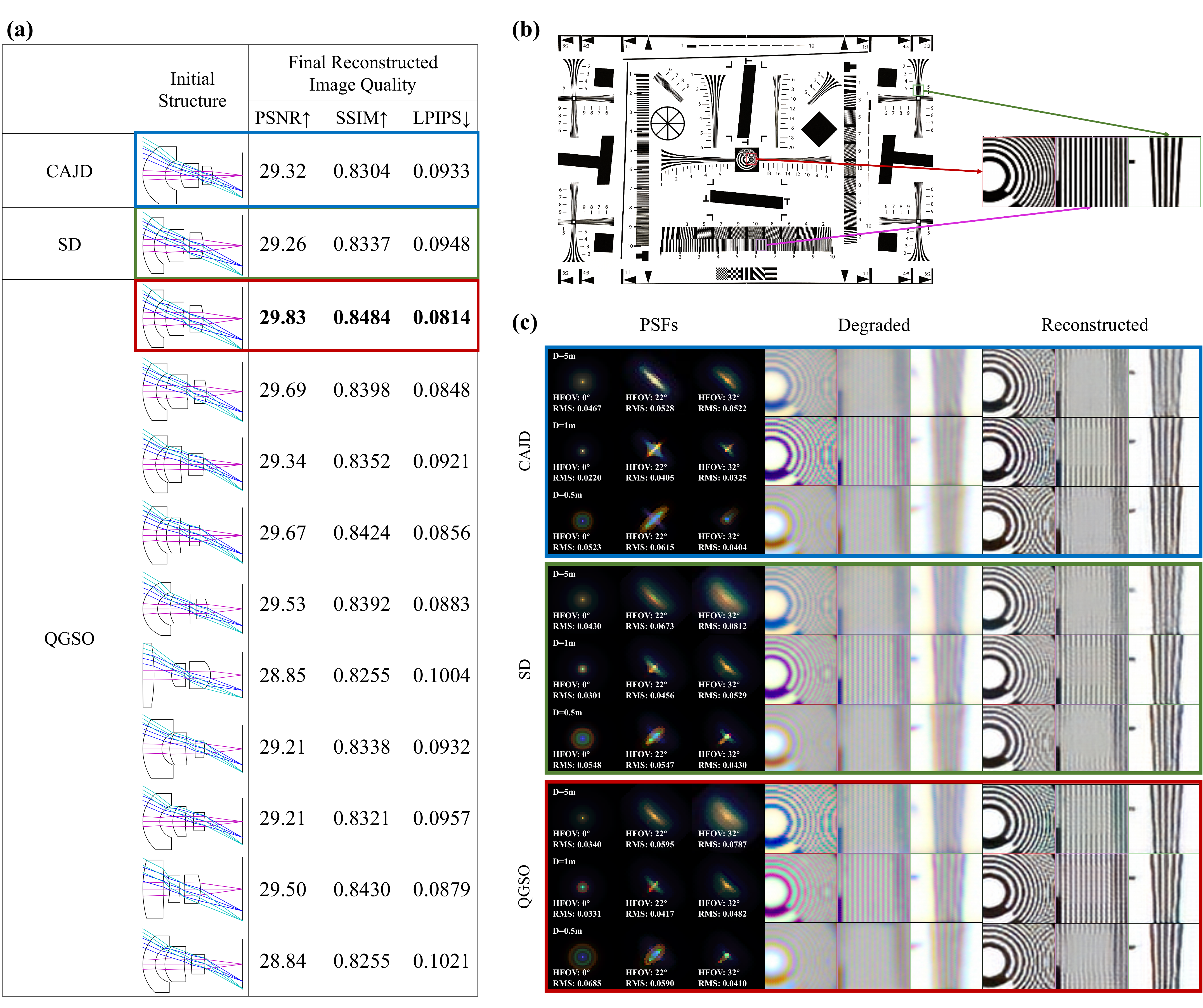}
  \vspace{-1.5em}
  \caption{Comparison between CAJD (CODE V Assisted Joint Design), SD (separate design), and QGSO under \textbf{3E-\uppercase\expandafter{\romannumeral2}}. (a) the initial structures and corresponding final reconstructed image quality of three methods. (b) the resolution chart (ISO 12233) taken by iPhone 12 and zoomed patches were used to evaluate image quality. (c) for each method and from left to right, we show 1) PSFs and corresponding RMS size ($mm$) across $3$ \textbf{D}epths (top: D${=}5m$; middle: D${=}1m$; bottom: D${=}0.5m$) and $3$ HFOV (left: $0^\circ$; middle: $22^\circ$; right: $32^\circ$); 2) degraded zoomed patches (top: D${=}5m$; middle: D${=}1m$; bottom: D${=}0.5m$); and 3) reconstructed zoomed patches (top: D${=}5m$; middle: D${=}1m$; bottom: D${=}0.5m$).}
  \label{fig:psf2}
  \vspace{-1.0em}
\end{figure*}

\begin{figure*}[!t]
  \centering
  \includegraphics[width=0.9\linewidth]{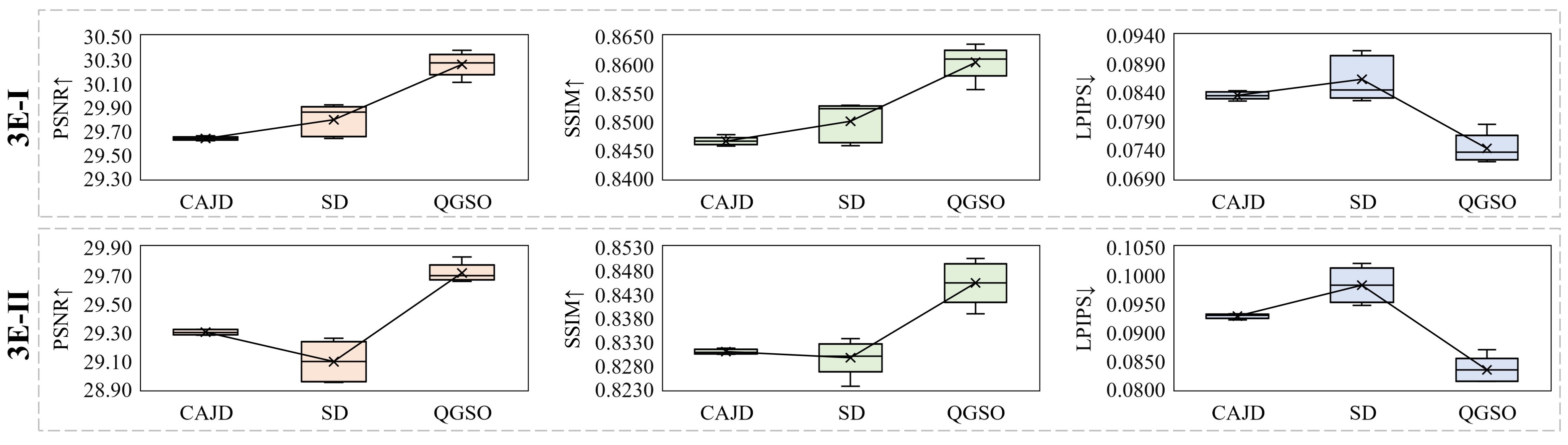}
  \vspace{-1.0em}
  \caption{Quantitative comparison between CAJD (CODE V Assisted Joint Design), SD (separate design), and QGSO under \textbf{3E-\uppercase\expandafter{\romannumeral1}} and \textbf{3E-\uppercase\expandafter{\romannumeral2}}. We use PSNR, SSIM, and LPIPS as evaluation metrics which are displayed from left to right.}
  \label{fig:final}
  \vspace{-1.5em}
\end{figure*}

After EPJO completes the joint optimization, the final solution of CAJD can be directly obtained because there is only $1$ initial structure. Differently, QGSO can obtain multiple final solutions, and we evaluate all the solutions using PSNR, SSIM~\cite{wang2004image}, and LPIPS~\cite{zhang2018unreasonable}. The final ranking basis is determined by averaging the rankings obtained from these metrics and we select the optimal result as the final solution of QGSO. 

Additionally, while the joint design method can theoretically explore the solution space more comprehensively compared to the Separate Design (SD) method due to its ability to synchronously optimize the optical system and image reconstruction model, the two methods have always lacked fair experiments for quantitative comparison. 
Therefore, we also design experiments to investigate the benefit of joint design methods in improving the upper limit of computational imaging system performance. 
To ensure the fairness of the experiment, the initial structure is consistent with the best initial structure searched by QGSO and the loss function is also set to Eq.~(\ref{eq:loss4}). The difference lies in that SD replaces the reconstructed image in Eq.~(\ref{eq:loss4}) with the degraded image for optimization and then fixes the designed optical system before training the reconstruction network. In other words, the optical system is independently designed without the reconstruction network, and then the reconstruction network is independently optimized. In addition, other training strategies of SD are consistent with QGSO, ensuring that the only factor affecting the final result is whether the optical system is co-designed with the reconstruction network so that we can test the benefit of joint design methods in improving the performance of the computational imaging system. 

\PAR{Experimental Results.} 
Finally, we obtain experimental results for CODE V Assisted Joint Design (CAJD) method, separate design (SD) method, and QGSO under two design specifications, 3E-\uppercase\expandafter{\romannumeral1} and 3E-\uppercase\expandafter{\romannumeral2}. Fig.~\ref{fig:psf1} demonstrates experimental results under 3E-\uppercase\expandafter{\romannumeral1}, and Fig.~\ref{fig:psf2} demonstrates experimental results under 3E-\uppercase\expandafter{\romannumeral2}. As shown in Fig.~\ref{fig:psf1}(a), under 3E-\uppercase\expandafter{\romannumeral1}, CAJD chooses the classic Cooke Triplet as the initial design, which is also one of QGSO's initial designs. However, QGSO searches for another structure from all $10$ results, which achieves improvements in PSNR/SSIM/LPIPS of $0.74dB/0.0170/0.0114$ compared to CAJD. In addition, even if SD uses the best initial structure found by QGSO to optimize the optical system and the reconstruction model separately, QGSO achieves improvements in PSNR/SSIM/LPIPS of $0.74dB/0.0168/0.0177$ compared to SD. Similarly, Fig.~\ref{fig:psf2}(a) shows that under 3E-\uppercase\expandafter{\romannumeral2}, QGSO achieves improvements in PSNR/SSIM/LPIPS of $0.51dB/0.0180/0.0119$ compared to CAJD and of $0.57dB/0.0147/0.0134$ compared to SD. Apart from the improvement in PSNR/SSIM/LPIPS, Fig.~\ref{fig:psf1}(c) and Fig.~\ref{fig:psf2}(c) shows the PSFs of the optical system, degraded images, and reconstructed images provided by three methods, which indicate that the imaging quality of the computational imaging system designed by QGSO is better at most depths and fields of view. 

Furthermore, we explore the reasons why QGSO can improve computational imaging quality by analyzing the characteristics of PSFs. 

\PAR{CAJD and QGSO.} 
Compared to CAJD, QGSO traverses more possible optical systems through OptiFusion, making it more likely to find structures that are more suitable for image reconstruction. Fig.~\ref{fig:psf2}(c) shows that under 3E-\uppercase\expandafter{\romannumeral2}, the characteristics of PSFs are quite different between CAJD and QGSO. The average spot RMS size of CAJD ($0.0445mm$) is even smaller than that of QGSO ($0.0515mm$). However, from the perspective of degraded and reconstructed images, QGSO's PSFs introduce a haze effect in the degraded images while effectively preserving image features~\cite{yang2023image}. In contrast, CAJD has smaller PSFs but blends the information. In addition, Fig.~\ref{fig:psf1}(c) shows that under 3E-\uppercase\expandafter{\romannumeral1}, the spot RMS size of CAJD may be even smaller at certain fields of view and \textbf{D}epths. 
For example, the spot RMS size of CAJD is $0.0534mm$ and the spot RMS size of QGSO is $0.0550mm$ when D${=}5m$ and HFOV${=}14^\circ$. From the perspective of degraded images and reconstructed images, however, the PSFs of CAJD significantly increase the loss of texture information, which leads to poorer image quality after the final reconstruction. Existing works generally use spot RMS size to measure the ability of PSFs to retain information~\cite{cote2023differentiable, yang2023image}. However, the experimental results indicate that PSFs with similar size but different aberration characteristics may also have significant differences in their ability to retain information, resulting in differences in the quality of reconstructed images. Therefore, the PSFs of the initial lens determined based on traditional optical design experience may not necessarily have the strongest ability to retain information. Differently, QGSO can automatically traverse diverse initial designs with different characteristics of PSFs, which can effectively avoid missing a more suitable initial lens for the reconstruction model.

\PAR{SD and QGSO.}
It can be observed from both Fig.~\ref{fig:psf1}(c) and Fig.~\ref{fig:psf2}(c) that the characteristics of PSFs are similar between SD and QGSO because the initial structures are consistent. Fig.~\ref{fig:psf1}(c) shows the average spot RMS size of SD ($0.0557mm$) is close to that of QGSO ($0.0550mm$), and Fig.~\ref{fig:psf2}(c) also shows the average spot RMS size of SD ($0.0525mm$) is close to that of QGSO ($0.0515mm$). The difference is that QGSO can better balance the spot size at different fields of view and \textbf{D}epths. 
For example, Fig.~\ref{fig:psf1}(c) shows that spot RMS size of SD is smaller when HFOV${=}0^\circ$ and larger when HFOV${=}20^\circ$ across all $3$ \textbf{D}epths, but the comprehensive quality of images reconstructed by QGSO is significantly better, which means that QGSO sacrifices a portion of the imaging quality of the central field of view to improve the imaging quality of the edge field of view, thereby maximizing the preservation of information in both the central and edge fields of view. 
Similarly, there is also such a phenomenon in Fig.~\ref{fig:psf2}(c). The spot RMS size of SD is smaller when HFOV${=}0^\circ$ and D${=}5m$ and larger when HFOV${=}32^\circ$ and D${=}5m$, but the comprehensive quality of images reconstructed by QGSO is still significantly better. Therefore, the main advantage of joint design over separate design is that it allows the optical system to more accurately weigh the PSFs of each field of view and \textbf{D}epth based on the preferences of the reconstruction model, resulting in better final reconstruction quality. 
In contrast, SD first has to fix the optical system and then optimize the reconstruction model, which may prevent the optical system parameters from being fine-tuned according to the needs of the reconstruction model and may not achieve the best cooperation of the two components.

\PAR{Statistical Analysis.}
We repeat this experiment $5$ times. As shown in Fig.~\ref{fig:final}, the standard deviation of QGSO's $5$ experimental results is slightly larger than that of CAJD because there exists a mechanism of random search in QGSO. Nevertheless, QGSO achieves improvements in PSNR/SSIM/LPIPS compared to CAJD in all $5$ experiments, which provides consistent results with the previous single experiment and proves this global random search mechanism increases the probability of finding a more suitable initial structure. Moreover, even if SD uses the same initial structures as QGSO, the results of all $5$ experiments have proven that QGSO can achieve better cooperation of the optical system and reconstruction model.

Overall, the reasons why QGSO can improve the final imaging performance of computational imaging systems can be summarized as follows:
\begin{itemize}
    \item QGSO can automatically traverse diverse initial designs with different characteristics of PSFs, rather than designing an initial structure for best aberration correction based on traditional optical design experience. This can effectively avoid missing the most suitable initial structure for the reconstruction model.
    \item QGSO includes EPJO, which can jointly optimize the optical system and reconstruction model, allowing the parameters of the optical system to be automatically fine-tuned to a better state according to the preferences of the reconstruction model.
\end{itemize}

\section{Conclusion}
We have introduced the QGSO, an end-to-end design framework capable of autonomously exploring solutions for compound lens based computational imaging systems. We demonstrate that as a crucial component of QGSO, OptiFusion can traverse diverse and reasonable initial designs compared to existing methods in multiple design forms such as Cooke Triplets or Double Gauss lenses. We also demonstrate the benefits of QGSO in improving the final performance of compound lens based computational imaging systems through the end-to-end design of EDoF three-element lenses and reveal the reasons why QGSO can improve the final reconstructed image quality.

It must be stressed, however, that although QGSO can design lenses that meet many manufacturing constraints, it has not completely solved the problem of lens manufacturing. As a complex engineering problem, the optical design also requires tolerance analysis, stray light analysis, consideration of lens assembly, and so on. In the future, QGSO can be integrated with existing commercial software, and the solutions found by QGSO can be directly input into existing commercial software for subsequent analysis.

There are some empirical hyper-parameters in this paper, most of which are the weights of specified physical quantities. The fundamental guideline is that these weights are roughly set based on the units and sizes of the corresponding physical quantities to ensure balance between different physical quantities. 
For example, the weight in Eq.~(\ref{eq:OptiFusion_loss_2}) that constrains TTL (Total Track Length) is set to $0.01$ and that constrains distortion is set to $1$ in the experiment because deviation of TTL from the constraint boundary by $1mm$ is roughly equivalent to the deviation of distortion from the constraint boundary by $1\%$. 
Although the experimental results have demonstrated the effectiveness of these hyper-parameters to some extent, further exploration of more optimal settings can be conducted in the future.

Looking ahead, OptiFusion can be combined with LensNet. Specifically, the development of a more comprehensive lens library through OptiFusion could serve as a means to train network models like LensNet for lens generation, accelerating the inference of suitable initial structures. 
Moreover, the post-processing algorithm used in QGSO can be replaced easily with any other visual task model according to specific needs in future research, and the comprehensive lens library established by OptiFusion could also be combined with specific visual task model to facilitate the analysis of visual task model preferences, substantially narrowing the range of initial structures to be screened.
This enhancement would significantly improve design efficiency because there is no need to traverse all possible initial structures if we have a sufficient understanding of visual task model preferences.

{\small
\bibliographystyle{IEEEtran}
\bibliography{bib}
}

\clearpage
\appendices
\counterwithin{figure}{section}
\counterwithin{equation}{section}

\section{Detailed Definitions of $\mathcal{L}_S$ and $\mathcal{L}_{LC}$}
As stated in Sec. III-A of the main text, we integrate a spot loss ($\mathcal{L}_S$) and a lateral chromatic aberration loss ($\mathcal{L}_{LC}$) to assess the imaging quality of an optical system in OptiFusion.
Here, $\mathcal{L}_S$ quantifies the average spot RMS radius~\cite{Foreman:74} across all sampled fields of view and wavelengths 
\begin{equation}
\label{eq:ls}
    \small \mathcal{L}_S=
    \frac{1}{n_{f} n_{w}} {\sum_{i=1}^{n_{f}}} {\sum_{j=1}^{n_{w}}} \sqrt{\frac{\sum_{k=1}^{n_{r}} (x(i,j;k)-\overline{x(i,j)})^2 + y(i,j;k)^2}{n_{r}}}.
\end{equation}
Here, $n_{f}$ represents number of sampled fields of view, $n_{w}$ represents number of sampled wavelengths, $n_{r}$ represents number of sampled rays, $x(i,j;k)$ represents the image plane $x$ coordinate of the $k_{th}$ ray, traced at $i_{th}$ sampled field of view and $j_{th}$ sampled wavelength, $y(i,j;k)$ represents the image plane $y$ coordinate of the $k_{th}$ ray, traced at $i_{th}$ sampled field of view and $j_{th}$ sampled wavelength, and $\overline{x(i,j)}$ represents the $x$ coordinate of the main ray at $i_{th}$ sampled field of view and $j_{th}$ sampled wavelength. We assume that the sampled object points are all in the $x$-axis direction so $\overline{y(i,j)} = 0$.

And $\mathcal{L}_{LC}$ accounts for the average lateral chromatic aberration~\cite{mahajan1991aberration} across all sampled fields of view
\begin{equation}
\begin{aligned}
\label{eq:lc}
    \mathcal{L}_{LC}= \frac{1}{n_{f}}{\sum_{i=1}^{n_{f}}} (&\max \{ \overline{x(i,1)}, \overline{x(i,2)}, ..., \overline{x(i,n_{w})} \}-\\
    &\min \{\overline{x(i,1)}, \overline{x(i,2)}, ..., \overline{x(i,n_{w})} \}).
\end{aligned}
\end{equation}
Here, at each sampled field of view, $\mathcal{L}_{LC}$ quantifies maximum distance between main ray positions across all sampled wavelengths.

\begin{figure*}[h]
  \centering
  \includegraphics[width=0.90\linewidth]{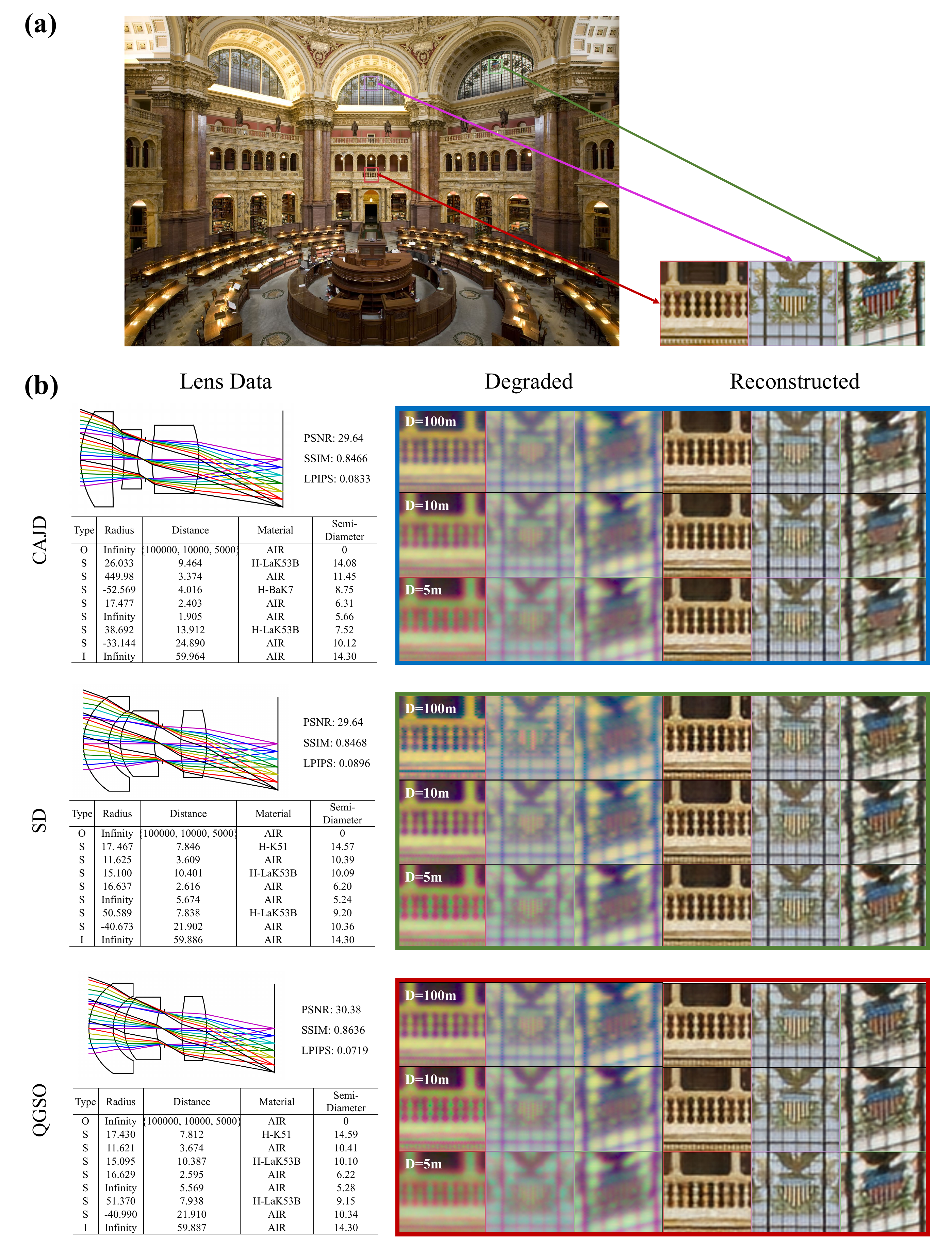}
  \vspace{-1.5em}
  \caption{Lens data and more visualization results for CAJD (CODE V Assisted Joint Design), SD (Separate Design), and QGSO under \textbf{3E-\uppercase\expandafter{\romannumeral1}}. (a) the clear image and zoomed patches used to evaluate image quality. (b) for each method and from left to right, we show 1) lens data; 2) degraded zoomed patches (top: D${=}100m$; middle: D${=}10m$; bottom: D${=}5m$); and 3) reconstructed zoomed patches (top: D${=}100m$; middle: D${=}10m$; bottom: D${=}5m$).}
  \label{fig:3p1}
\end{figure*}

\begin{figure*}[h]
  \centering
  \includegraphics[width=0.90\linewidth]{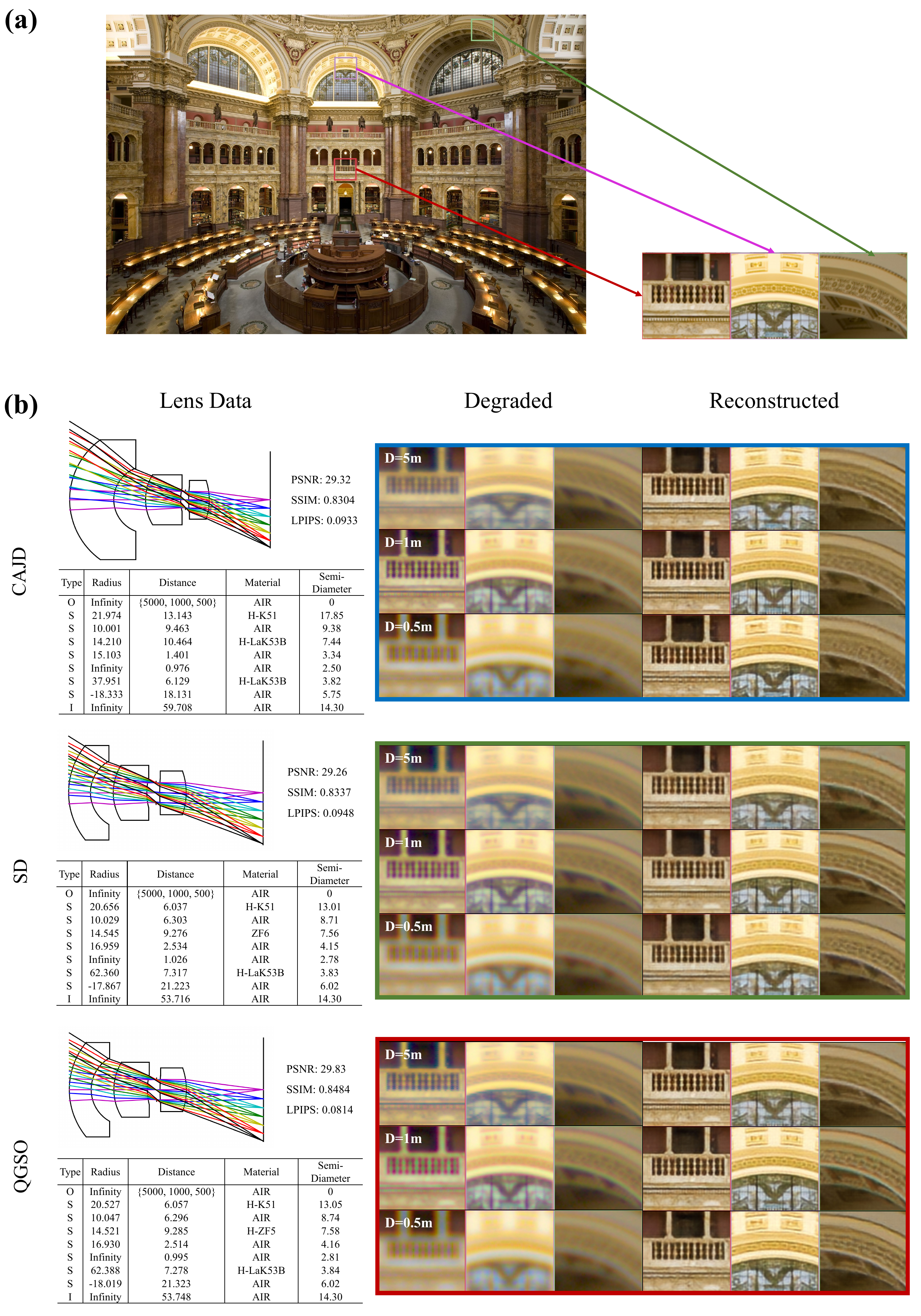}
  \vspace{-1.5em}
  \caption{Lens data and more visualization results for CAJD (CODE V Assisted Joint Design), SD (Separate Design), and QGSO under \textbf{3E-\uppercase\expandafter{\romannumeral2}}. (a) the clear image and zoomed patches used to evaluate image quality. (b) for each method and from left to right, we show 1) lens data; 2) degraded zoomed patches (top: D${=}5m$; middle: D${=}1m$; bottom: D${=}0.5m$); and 3) reconstructed zoomed patches (top: D${=}5m$; middle: D${=}1m$; bottom: D${=}0.5m$).}
  \label{fig:3p2}
\end{figure*}

\section{Comparison Experiment Between OptiFusion and LensNet} 
\subsection{Physical Constraints in Multiple Design Forms}
As stated in Sec. V-A of the main text, LensNet is confined to basic specifications including Effective Focal Length (EFL), F-number, and Half Field Of View (HFOV), without accommodating more complex physical constraints. As outlined in Table~\ref{tab:lesnet}, however, the physical constraints that can be considered by OptiFusion include distortion, glass center thickness, air center spacing, glass edge thickness, air edge spacing, BFL (Back Focal Length), TTL (Total Track Length), HFOV, EFL, F-number, curvature, refractive index, abbe number and so on. 

Therefore, to ensure a fair comparison, we first set a certain design specification with a $40mm$ EFL, an F-number of $2.5$, and an HFOV of $20^{\circ}$ and use LensNet to produce designs under this specification. 
After obtaining the output results of LensNet in multiple design forms, we apply OptiFusion to produce designs with reasonable physical constraints that refer to the physical constraints of lenses generated by LensNet, as outlined in Table~\ref{tab:lesnet}. Specifically, when a certain design form exhibits the following two situations, the relevant physical constraints are set based on optical design experience:

1) there are no matched structures in some design forms (GAGAGASAGAGA, SAGAGAGAGAGAGA, and GAGAGASAGAGAGA), so there are also no physical constraints that can be referenced, and physical constraints in these design forms are set based on optical design experience. 

2) Under certain design forms (GASGAGGA, GAGAGASAGA, and GAGGGSAGGA), the lenses output by LensNet may have surface overlap, which means that glass edge thickness or air edge spacing may be less than $0$. At this point, the minimum values of the glass edge thickness or air edge spacing are not taken as $0mm$, but a reasonable number (generally $\geq 0.5mm$) based on optical design experience.

\begin{table*}[!t]
    \begin{center}
        \caption{Physical constraints of multiple design forms.}
        \label{tab:lesnet}
        \resizebox{0.8\textwidth}{!}
{
\renewcommand{\arraystretch}{1.2}
\setlength{\tabcolsep}{1mm}{

\begin{tabular}{ccccccccc}
\hline
Design Form  & GAGASAGA  & GASGAGGA & GAGAGASAGA & GAGGGSAGGA & GAGGASAGGAGA & GAGAGASAGAGA & SAGAGAGAGAGAGA & GAGAGASAGAGAGA \\ \hline \hline
\multicolumn{1}{c|}{Distortion} & $-1\%\sim1\%$ & $-1\%\sim1\%$ & $-1\%\sim1\%$ & $-3.7\%\sim3.7\%$ & $-5\%\sim5\%$ & $-3\%\sim3\%$ & $-3\%\sim3\%$ & $-3\%\sim3\%$  \\
\multicolumn{1}{c|}{Glass center thickness} & $1.3mm\sim8mm$ & $1.8mm\sim6mm$ & $1.25mm\sim10mm$ & $2mm\sim10mm$ & $2mm\sim10mm$ & $1.5mm\sim6mm$ & $1.5mm\sim8mm$ & $1.5mm\sim6mm$\\
\multicolumn{1}{c|}{Air center spacing}     & $1.3mm\sim10mm$ & $2mm\sim5mm$ & $0.5mm\sim10mm$ 
& $0.4mm \sim 8mm$ & $0.5mm \sim 15mm$ & $0.1mm \sim 15mm$ & $0.1mm \sim 15mm$ & $0.1mm \sim 15mm$\\
\multicolumn{1}{c|}{Glass edge thickness}   & $1mm\sim8mm$ & $1mm\sim8mm$  & $1mm\sim10mm$ & $0.68mm\sim12.6mm$ & $0.8mm\sim10mm$ & $1.2mm\sim6mm$ & $1.2mm\sim8mm$ & $1.2mm\sim6mm$\\
\multicolumn{1}{c|}{Air edge spacing}       & $0.3mm\sim8mm$ & $1mm\sim8mm$ & $0.5mm\sim15mm$
& $0.5mm\sim8mm$ & $0.5mm\sim15mm$ & $0.5mm\sim15mm$ & $0.5mm\sim15mm$ & $0.5mm\sim15mm$\\ 
\multicolumn{1}{c|}{BFL}                    & $>29mm$ & $>30mm$ & $>24mm$ & $>30mm$ & $>24mm$ & $>26mm$ & $>18mm$ & $>26mm$\\
\multicolumn{1}{c|}{TTL}                    & $<50mm$ & $<50mm$ & $<47mm$ & $<46mm$ & $<64mm$ & $<56mm$ & $<80mm$ & $<65mm$\\
\hline

\multicolumn{1}{c|}{HFOV}                   & \multicolumn{8}{c}{$25^{\circ}$}  \\
\multicolumn{1}{c|}{EFL}                    & \multicolumn{8}{c}{$40mm$}  \\
\multicolumn{1}{c|}{F-number}               & \multicolumn{8}{c}{$2.5$}  \\
\multicolumn{1}{c|}{Curvature}              & \multicolumn{8}{c}{$-0.11\sim0.11$}  \\
\multicolumn{1}{c|}{Refractive index}       & \multicolumn{8}{c}{$1.51\sim1.76$}  \\
\multicolumn{1}{c|}{Abbe number}            & \multicolumn{8}{c}{$27.5\sim71.3$}  \\
                
\hline
\end{tabular}
}
}
    \end{center}
\end{table*}
 
\section{End-to-end Design of EDoF Three-element Lenses} 
\subsection{Detailed Lens Data And More Visualization Results}
We evaluate the global search capability of QGSO in Sec. V-B of the main text by comparing it with both the CAJD (CODE V assisted joint design) and the SD (Separate Design) under two design specifications, 3E-\uppercase\expandafter{\romannumeral1} and 3E-\uppercase\expandafter{\romannumeral2}. In addition to Fig. 5 and Fig. 6 in the main text, we also provide detailed lens data and more examples of image reconstruction in Fig.~\ref{fig:3p1} and Fig.~\ref{fig:3p2}. Firstly, detailed lens data indicates that the glass center thickness, air center spacing, BFL, TTL, \etc, all meet the physical constraints set in Table 1 of the main text, which demonstrates QGSO's consideration of manufacturing constraints. In addition, more examples of image reconstruction also demonstrate the superior performance of the computational imaging system designed by QGSO.

\end{document}